\documentclass[11pt,a4paper,twoside]{article}

\usepackage{axodraw}
\usepackage[latin1]{inputenc}
\def\be{\begin{equation}}
\def\ee{\end{equation}}
\def\ba{\begin{array}}
\def\bacc{\begin{array} {cc}}
\def\ea{\end{array}}
\def\bea{\begin{eqnarray}}
\def\eea{\end{eqnarray}}
\def\bd{\begin{displaymath}}
\def\ed{\end{displaymath}}
\def\ha{\hat{\alpha}}
\def\hb{\hat{\beta}}
\def\D{\mathcal{D}}
\def\a{\alpha}
\def\b{\beta}
\def\c{\gamma}
\def\d{\delta}
\def\o{\omega}
\def\mv{\mathcal{V}}
\def\e{\mbox{det}\left(e^{\a}_m\right)}
\def\U{\mathcal{U}}

\begin{document}
\begin{center}

{\Large\bf On the Decoupling of Heavy Modes in Kaluza-Klein
Theories}

\vspace{1cm}

{\large S. Randjbar-Daemi$^a$\footnote{Email: seif@ictp.trieste.it},
A. Salvio$^b$\footnote{Email: salvio@sissa.it} and M.
Shaposhnikov$^c$\footnote{Email:
Mikhail.Shaposhnikov@epfl.ch}}\\

\vspace{.6cm}

{\it {$^a$ International Center for Theoretical Physics, \\Strada
Costiera 11, 34014 Trieste, Italy}}

\vspace{.4cm}

{\it {$^b$ International School for Advanced Studies,\\
Via Beirut 2-4, 34014 Trieste, Italy}}

\vspace{.4cm}

{\it {$^c$ Institut de Th\'eorie des Ph\'enom\`enes Physiques,\\
  \'Ecole Polytechnique F\'ed\'erale de Lausanne,\\
  CH-1015 Lausanne, Switzerland}} \vspace{.4cm}

\end{center}

\vspace{1cm}

\begin{abstract}

In this paper we examine the 4-dimensional effective theory for the
light Kaluza-Klein (KK) modes.  Our main interest is in the
interaction terms. We point out that the  contribution of the heavy
KK modes is generally needed  in order to reproduce the correct
predictions for the observable quantities involving the light modes.
As an example we study  in some detail a 6-dimensional
Einstein-Maxwell theory coupled to a charged scalar and  fermions.
In this case the contribution of the heavy KK modes are
geometrically interpreted as the deformation of the internal space.

\end{abstract}

\newpage

\tableofcontents

\newpage

\section{Introduction} \label{intro}
\setcounter{equation}{0}

 In studying the low energy physics of the
light modes of a 4+d dimensional theory the attention is usually
paid only to the spectral aspects. After determining the quantum
numbers of the light modes  the nature and the form of the
interaction terms are often assumed to be dictated by symmetry
arguments.  Such arguments fix the general form of all the
renormalilzable terms and if the effective theory is supersymmetric
certain relationship between the couplings can also be established
by supersymmetry. The masses are derived from the bilinear part of
the effective action and the role of the heavy modes in the actual
values of the masses and the couplings of the effective theory for
the light modes are seldom taken into account. It is, however, well
known from the study of the GUT's in 4-dimensions that the heavy
modes have an important role to play even at low energies
\cite{GUT}. This happens through their contributions to the
couplings  entering into the effective Lagrangians describing the
low energy physics of the light modes. According to Wilsonian
approach, in order to obtain an effective theory applicable in large
distances, the heavy modes should be integrated out \cite{Wilson,
Weinberg:1980wa}. The processes of "integrating out" has the effect
of modifying the couplings of the light modes or introducing
additional terms which are suppressed by inverse powers of the heavy
masses \cite{Appelquist:1974tg}.

The aim of the present paper is to examine the role of the heavy
modes in the low energy description of a higher dimensional theory.
To this end we shall basically perform two complementary
calculations. The first one will start from a solution of a higher
dimensional theory with a 4-dimensional Poincar\'e invariance and
develop an action functional for the light modes of the effective
4-dimensional theory. This effective action generally has a local
symmetry which should be broken by  Higgs mechanism. Our interest is
in the spectrum of the broken theory. The procedure is essentially
what is adopted in the effective description of higher dimensional
theories including superstring and M-theory compactifications. In
this construction the heavy KK modes are generally ignored simply by
reasoning that their masses are of the order of the compactification
mass and this can be as heavy as the Planck mass. Therefore they can
not affect the low energy physics of the light modes.

 In the second approach which we shall call the $\it{ geometrical\ approach}$ we shall find a solution of the higher dimensional equations
with the same symmetry group as the one of the broken phase of the
effective 4-dimensional theory for the light modes. We shall then
study the physics of the 4-dimensional light modes around this
solution. The result for the effective 4-dimensional theory will
turn out to be different from the first approach.  The aim of this
paper is to show that the difference is precisely due to the fact
that in constructing the effective theory along the lines of the
first approach the contribution of the heavy KK modes have been
ignored. Indeed it will be argued - and demonstrated by working out
some explicit examples - that taking due care of the role of the
heavy modes a complete equivalence is established between the two
approaches.

To motivate the discussion in a simpler context, in section
\ref{simple4D} we shall work out a simple model of two coupled
scalar fields in 4-dimensions which will be generalized to a
multiplet of scalar fields in arbitrary dimensions in section
\ref{general}. The examples in sections \ref{simple4D} and
\ref{general} will clarify the relevance of the heavy modes in the
low energy description of the light modes. In sections \ref{def} and
\ref{u13} we shall study a higher dimensional (in this case six
dimensional) theory of Einstein-Maxwell system \cite{RSS} coupled to
a charged scalar and eventually also to charged fermions. Such a
model can arise in the compactification of string or M-theory to
lower dimensions.  The system has enough number of adjustable
parameters to allow us to go to various limits in order to establish
the main point of our paper. The result will of course confirm the
above mentioned expectation that in order to obtain a correct
4-dimensional description of the physics of the light modes the
contribution of the heavy modes should be duly taken into
account\footnote{Of course this doesn't prove that the heavy modes
contribution never vanishes: for instance \cite{Gibbons:2003gp}
proves the decoupling of the heavy modes in the $(Minkowski)_4\times
S^2$ compactification of the 6-dimensional chiral supergravity
\cite{Salam:1984cj}, which is basically the supersymmetric version
of our 6-dimensional theory.}. This example is particularly
interesting because the first kind of solution will produce an
effective 4-dimensional gauge theory with a $SU(2)\times U(1)$
symmetry which will be broken to $U(1)$ by a complex triplet of
Higgs fields. The geometrical approach, on the other hand, will take
us directly to the unbroken $U(1)$ phase by deforming a round sphere
into an ellipsoid\footnote{ This corresponds to the magnetic
monopole charge of 2 as explained in section 5.  A monopole charge
of unity will produce a Higgs doublet of SU(2). }. In the
geometrical approach the W and the Z masses originate from the
deformation of the internal space. In this sense the standard Higgs
mechanism acquires a geometrical origin\footnote{It should be
mentioned that all of our discussion is ( semi-) classical.  To
include quantum and renormalization effects is beyond the scope of
the present paper.}. We elaborate a little more on this point in
section \ref{conclusions} which summarizes our results.  Some
technical aspects of various derivations have been detailed in the
appendices.

\section{A Simple 4D Theory} \label{simple4D}
\setcounter{equation}{0}

 Let's consider a 4-dimensional theory,
which contains two real scalar fields $\varphi$ and $\chi$ and with
the lagrangian
$$ \mathcal{L}=
-\frac{1}{2}\partial_{\mu}\varphi\partial^{\mu}\varphi-\frac{1}{2}\partial_{\mu}\chi\partial^{\mu}\chi
-\frac{1}{2}m^2_{\varphi}\varphi^2-\frac{1}{2}m^2\chi^2-\frac{1}{4}\lambda_{\varphi}\varphi^4
-\frac{1}{4}\lambda_{\chi}\chi^4-a\varphi^2\chi^2, $$
where $m^2_{\varphi}$, $m^2$, $\lambda_{\varphi}$, $\lambda_{\chi}$
and $a$ are real parameters\footnote{Of course we consider only the
values of these parameters such that the scalar potential is bounded
from below.}. Here we have the symmetry:
\bea &&Z_2: \varphi\rightarrow \pm \varphi, \nonumber \\
&&Z_2': \chi \rightarrow \pm \chi. \eea
 This is a very particular
example and of course we don't want to present any general result in this section, we just want to
provide a framework in which the general equivalence that we spoke about in the introduction emerges in a simple
way and is not obscured by technical difficulties.

For $m^2_{\varphi}<0$ we have the following solution of the
equations of motion (EOM):
\be \chi=0, \quad \quad
\varphi=\sqrt{\frac{-m^2_{\varphi}}{\lambda_{\varphi}}}\equiv
\varphi_{eff}, \label{simplefirstbac}\ee
which breaks $Z_2$ but preserves $Z_2'$. We can express the
lagrangian in terms of the fluctuation $\d \varphi $ and $\chi$
around this background:
 \bea \mathcal{L}&=&
-\frac{1}{2}\partial_{\nu}\d\varphi\partial^{\nu}\d\varphi-\frac{1}{2}\partial_{\nu}\chi\partial^{\nu}\chi
+m^2_{\varphi}\left(\d\varphi\right)^2-\frac{1}{2}\mu^2\chi^2
\nonumber
\\
&&-\sqrt{-m^2_{\varphi}\lambda_{\varphi}}\left(\d\varphi\right)^3-\frac{1}{4}\lambda_{\varphi}\left(\d\varphi\right)^4
-\frac{1}{4}\lambda_{\chi}\chi^4-2a\sqrt{\frac{-m^2_{\varphi}}{\lambda_{\varphi}}}\d\varphi\chi^2
\nonumber \\&&
-a\left(\d\varphi\right)^2\chi^2+constants,\label{lagexp} \eea
where
\be \mu^2\equiv m^2-2a\frac{m^2_{\varphi}}{\lambda_{\varphi}}. \ee

\begin{figure}[t]
\begin{center}
\begin{picture}(300,100)(0,0)

\DashLine(50,80)(190,80){4}\Text(120,100)[]{Heavy scalar}
\Line(190,80)(230,50)\Text(210,55)[]{$\chi$}
\Line(190,80)(230,110)\Text(210,115)[]{$\chi$}
\Line(10,50)(50,80)\Text(30,55)[]{$\chi$}
\Line(10,110)(50,80)\Text(30,115)[]{$\chi$}

\end{picture}
\end{center}
\caption{\footnotesize A tree diagram which describes the scattering
of two light $\chi$, through the exchange of an heavy scalar. This
kind of diagram gives a contribution to the quartic term in the
effective theory potential.} \label{simplediagram}
\end{figure}
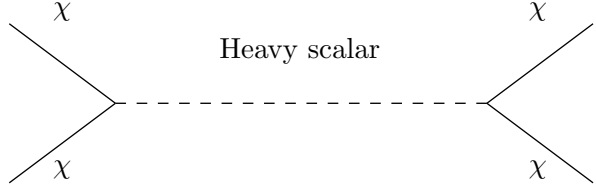

If $|\mu^2|\ll |m^2_{\varphi}|$, we expect that the heavy mode
$\d\varphi$ can be integrated out and an effective theory for
$\chi$ can be constructed for both the signs of $\mu^2$. However
it's important to note that $\d\varphi$ cannot be simply neglected
because it gives a contribution, because of the
trilinear\footnote{Also the quartic coupling
$\left(\d\varphi\right)^2\chi^2$ gives a contribution to the
operator $\chi^4$, but this is negligible in the classical limit.}
coupling $\d\varphi \chi^2$ in (\ref{lagexp}), to the operator
$\chi^4$ in the effective theory, through the diagram
\ref{simplediagram}. This is similar to what is usually done in GUT theories \cite{GUT}, where, for instance, four fermions
effective interactions emerge by integrating out the heavy gauge fields \cite{Buras:1977yy}.
At the classical level the effective lagrangian for $\chi$ is
\be
\mathcal{L}_{eff}=-\frac{1}{2}\partial_{\nu}\chi\partial^{\nu}\chi-\frac{1}{2}\mu^2\chi^2
-\frac{1}{4}\left(\lambda_{\chi}-\frac{4a^2}{\lambda_{\varphi}}\right)\chi^4+...\,,\label{effsimple}
\ee
where the dots represent higher dimensional operators. The term
$a^2\chi^4/\lambda_{\varphi}$ is the contribution of the heavy
mode. The result (\ref{effsimple}) was originally derived in \cite{Chan:1979ce}, but here we
want also to study the effective theory with spontaneous symmetry breaking and we want to compare it with the
low energy limit of the fundamental theory.

 For $\mu^2>0$, the minimum of the effective theory potential
is for $\chi=0$. Instead for $\mu^2<0$ we have
\be
\chi=\sqrt{\frac{-\mu^2}{\lambda_{\chi}-\frac{4a^2}{\lambda_{\varphi}}}}\label{chieff}\ee
and the fluctuation $\d\chi$ over this background has the
following mass squared:
\be M^2(\d\chi)=-2\mu^2. \label{Meff}\ee
This results will be not modified by the higher dimensional operator
at the leading order\footnote{The mass $\mu$ is small in the sense
$|\mu|\ll|m_{\varphi}|$.} in $\mu$. The equations (\ref{chieff}) and
(\ref{Meff}) represent the effective theory prediction for the VEV
and the spectrum in the phase where $Z_2'$ is broken.

On the other hand, a solution of the fundamental EOM, namely the EOM
derived from the fundamental lagrangian $\mathcal{L}$, is
\bea
\chi^2&=&\frac{-\mu^2}{\lambda_{\chi}-\frac{4a^2}{\lambda_{\varphi}}}+O(\mu^3),\nonumber
\\
\varphi^2&=&-\frac{m^2_{\varphi}}{\lambda_{\varphi}}+\frac{2a\mu^2}{\lambda_{\varphi}\lambda_{\chi}-4a^2}+O(\mu^3)
\label{correctsimpleVEV}\eea
which is a small deformation of (\ref{simplefirstbac}) at the
leading non trivial order in $\mu$ and breaks the $Z_2'$ symmetry.
Moreover the light mode which corresponds to this solution has a
mass squared $-2\mu^2$.

Therefore the effective theory prediction for the light mode VEV and
spectrum is correct, at the order $\mu$, in this simple framework,
but the heavy mode contribution is necessary in order the effective
theory prediction to be correct.

\section{A More General Case}\label{general}
\setcounter{equation}{0}

Now we want to extend the result of section \ref{simple4D} and ref
\cite{Chan:1979ce} to a more general class of theories. We consider
a set of real D-dimensional scalars $\Phi_i$ with a general
potential $V$: the lagrangian is

\be \mathcal{L}=-\frac{1}{2}\partial_M\Phi_i\partial^M\Phi_i -
V(\Phi), \ee
where $M,N,...$ run over all the space-time dimensions, while
$\mu,\nu,...$ and $m,n,...$ are respectively the 4-dimensional and
the internal coordinates indices. The EOM are
\be
\partial_M\partial^M\Phi_i-\frac{\partial V}{\partial\Phi_i}(\Phi)=0.\label{generalEOM}
\ee
We consider now a solution $\Phi_{eff}$ of (\ref{generalEOM})
which preserves the 4-dimensional Poincaré invariance and some
internal symmetry group $\mathcal{G}$; the corresponding mass squared eigenvalue
problem for the 4-dimensional states is
\be -\partial_m\partial_m\d \Phi_i+\frac{\partial^2V}{\partial\Phi_i\partial
\Phi_j}(\Phi_{eff})\d \Phi_j=M^2\d\Phi_i,\label{generaleigen}
\ee
where $\d \Phi$ is the fluctuation around $\Phi_{eff}$. We assume
that there are $n$ normalizable solutions $\D_l$ with small
eigenvalues ($M^2\sim \mu^2$), other, in principle infinite,
solutions\footnote{In principle $h$ can be a discrete or a
continuous variable.} $\tilde{\D}_h$with large eigenvalues ($M^2\gg
|\mu^2|$) and nothing else. These hypothesis are needed in order to
define the concept of light KK modes.

We can expand the scalars $\Phi_i$
as follows
\be
\Phi_i=\left(\Phi_{eff}\right)_i+\chi_l(x)\D_{li}(y)+\tilde{\chi}_h(x)\tilde{\D}_{hi}(y),\label{expansion}
\ee
where $\chi_l$ and $\tilde{\chi}_h$ are respectively the light and
heavy KK modes. We choose the $\D_l$ and $\tilde{\D}_h$ in order
that they form an orthonormal basis for the functions over the
internal space:
\bea \left<\D_l|\D_{l'}\right>&\equiv& \int
d^{D-4}y\D_{li}(y)\D_{l'i}(y)=\d_{ll'},\nonumber \\
\left<\tilde{\D}_h|\tilde{\D}_{h'}\right>&\equiv& \int
d^{D-4}y\tilde{\D}_{hi}(y)\tilde{\D}_{h'i}(y)=\d_{hh'},\nonumber
\\
\left<\D_l|\tilde{\D}_{h}\right>&\equiv& \int
d^{D-4}y\D_{li}(y)\tilde{\D}_{hi}(y)=0. \eea
We note that $\chi_l$ and $\tilde{\chi}_h$ could both belong to some
non trivial representation of the internal symmetry group
$\mathcal{G}$.

\subsection{The Effective Theory Method}\label{EFFGeneral}

We construct now some relevant terms in the effective theory for the
light KK modes $\chi_l$. Here "relevant terms" mean relevant terms
in the classical limit and in case we have a small point of minimum
of the order $\mu$ of the effective theory potential: we want to
compare the results of the effective theory for the light KK modes
with the low energy limit of the fundamental theory expanded around
a vacuum which is a small perturbation of $\Phi_{eff}$. Further we
calculate everything at leading non trivial order\footnote{The $\mu$
mass scale is small in the sense $|\mu|$ is much smaller than the
heavy masses.} in $\mu$.
 The relevant terms can be
computed by putting just the light KK modes in the action and
performing the integration over the extra dimensions and then by
taking into account the effect of heavy KK modes through the
diagrams like figure \ref{simplediagram}. In order to calculate
those diagrams, we give the interactions between two light modes
$\chi_l$ and one heavy mode $\tilde{\chi}_h$:
\be -\frac{1}{2}\left(\int d^{D-4}y
V_{ijk}\D_{li}\D_{mj}\tilde{\D}_{hk}\right)\chi_l\chi_m\tilde{\chi}_h,
\ee
where we have used the notation
\be V_{i_1 ... i_N}\equiv \frac{\partial^N V}{\partial\Phi_{i_1}
... \partial\Phi_{i_N} }\left(\Phi_{eff}\right). \ee
We get the following relevant terms in the effective theory
potential $\U$:
\be \U(\chi)=\frac{1}{2}c_l\mu^2\chi_l\chi_l
+\frac{1}{3}\lambda^{(3)}_{lmp}\chi_l\chi_m\chi_p+\frac{1}{4}\lambda_{lmpq}^{(4)}\chi_l\chi_m\chi_p\chi_q+...
\,,\label{Ugeneral}\ee
where the dots represent non relevant terms, $c_l$ are
dimensionless numbers and
\bea\lambda^{(3)}_{lmp}&\equiv&\frac{1}{2}\int d^{D-4}y
V_{ijk}\D_{li}\D_{mj}\D_{pk},\\
\lambda_{lmpq}^{(4)}&\equiv&\frac{1}{3!}\left(\int d^{D-4}y
V_{ijkk'}\D_{li}\D_{mj}\D_{pk}\D_{qk'}\right)+a_{lmpq},
\label{l4}\eea
where the quantities $a_{lmpq}$ represent the heavy modes
contribution and they are given by
\be a_{lmpq}=c_{lmpq}+c_{lpmq}+c_{lqpm} \ee
and
\be c_{lmpq}\equiv -\frac{1}{6}\int
d^{D-4}yd^{D-4}y'V_{ijk}(y)\D_{li}(y)\D_{mj}(y)
G_{kk'}(y,y')V_{i'j'k'}(y')\D_{pi'}(y')\D_{qj'}(y').\ee
The object $G_{kk'}$ is the Green's function for the mass squared operator at
the left hand side of (\ref{generaleigen}) and it's explicitly
given by
\be G_{kk'}(y,y')=\sum_h
\frac{1}{m_h^2}\tilde{\D}_{hk}(y)\tilde{\D}_{hk'}(y'),\ee
where $m^2_h$ is the eigenvalue associated to the eigenfunction
$\tilde{\D}_h$.

In the rest of this section we consider the predictions of the
effective theory with spontaneous symmetry breaking.  The potential
(\ref{Ugeneral}) has to be considered as a generalization of
(\ref{effsimple}), which was originally derived in
\cite{Chan:1979ce}. A non vanishing VEV breaks in general
$\mathcal{G}$ to some subgroup and it must satisfies
\bea \frac{\partial
\U}{\partial\chi_l}&=&c_l\mu^2\chi_l+\lambda^{(3)}_{lmp}\chi_m\chi_p+\lambda_{lmpq}^{(4)}\chi_m\chi_p\chi_q=
0. \label{effvacua}\eea
Since we require that $\chi_l$ goes to zero as $\mu$ goes to zero we
have
\be \chi_l=\chi_{l1}+\chi_{l2}+... \ee
where $\chi_{l1}$ is proportional to $\mu$, $\chi_{l2}$ is
proportional to $\mu^2$ and so on. At the order $\mu^2$ the
equations (\ref{effvacua}) reduce to
\be \lambda^{(3)}_{lmp}\chi_{m1}\chi_{p1}=0 \ee
which implies
\be \lambda^{(3)}_{lmp}\chi_{p1}=0. \label{eff2vacua}\ee
While, at the order $\mu^3$, the equations (\ref{effvacua}) reduce
to
\be
c_l\mu^2\chi_{l1}+\lambda_{lmpq}^{(4)}\chi_{m1}\chi_{p1}\chi_{q1}=0,\label{eff3vacua}
\ee
where we have used the equations (\ref{eff2vacua}).

Finally the mass spectrum corresponding to a solution of
(\ref{effvacua}) is given by the eigenvalues of the hessian matrix
of $\U$ in that solution:
\be \frac{\partial^2 \U}{\partial\chi_l
\partial
\chi_{l'}}=c_l\mu^2\d_{ll'}+2\lambda^{(3)}_{ll'm}\chi_m+
3\lambda_{ll'mq}^{(4)}\chi_m\chi_q. \ee
If we assume, for simplicity, $\lambda^{(3)}_{ll'm}=0$, which
corresponds to the absence of cubic terms in $\U$, the leading
order approximation of the hessian is simply given by
\be \frac{\partial^2 \U}{\partial\chi_l
\partial
\chi_{l'}}=c_l\mu^2\d_{ll'}+
3\lambda_{ll'mq}^{(4)}\chi_{m1}\chi_{q1}+O(\mu^3).\label{hessian}
\ee
In subsection \ref{correct} we show that this matrix, which
represents the mass spectrum for the light KK modes, and the
equations (\ref{eff2vacua}) and (\ref{eff3vacua}) for the light
modes VEVs are exactly reproduced by a D-dimensional analysis.

\subsection{D-dimensional analysis}\label{correct}

Now we want to find a solution of (\ref{generalEOM}) which is a
small perturbation, of the order $\mu$, of $\Phi_{eff}$ and then we
want to find the low energy mass spectrum of the fluctuations around
this solution. In general this solution will break $\mathcal{G}$ to
some subgroup like a solution of (\ref{effvacua}) does in the
effective theory method. The explicit form of such solution in the
simple case of section \ref{simple4D} is given by
(\ref{correctsimpleVEV}) and the low energy mass spectrum in that
simple case is represented by the squared mass $-2\mu^2$; now we
want to generalize these results.

Let's consider the expansion (\ref{expansion}); we observe that the
statement that the solution is a small perturbation of $\Phi_{eff}$
means
\bea&&\chi_l=\chi_{l1}+\chi_{l2}+...\,, \nonumber\\
&&\tilde{\chi}_h=\tilde{\chi}_{h1}+\tilde{\chi}_{h2}+...\,,\label{analitic}
\eea
that is there are no big $\mu-$independent terms in $\chi_l$ and
$\tilde{\chi}_h$. We consider now a Taylor expansion of the
equations (\ref{generalEOM}) around $\Phi_{eff}$:
\bea
&&\partial_m\partial_m\left(\Phi_i-\left(\Phi_{eff}\right)_i\right)\nonumber
\\
&&-\sum_{k=1}^{N}\frac{1}{k!}V_{ii_1...i_k}
\left(\Phi_{i_1}-\left(\Phi_{eff}\right)_{i_1}\right)\cdot ... \cdot
\left(\Phi_{i_k}-\left(\Phi_{eff}\right)_{i_k}\right)\nonumber\\
&&+O(\mu^{N+1})=0.\label{pertEOM}\eea
At the order $\mu$ the equations (\ref{pertEOM}) reduce to
\be \left(\partial_m
\partial_m\d_{ij}-V_{ij}\right)\left(\Phi_{j}-\left(\Phi_{eff}\right)_{j}\right)+O(\mu^2)=0,
\ee
which simply states
\be \tilde{\chi}_{h1}=0.\label{heavyVEV}\ee
 Moreover at the order $\mu^2$ the
equations (\ref{pertEOM}) imply
\be
\tilde{\chi}_{h2}\left(\partial_m\partial_m\d_{ij}-V_{ij}\right)\tilde{\D}_{hj}=
\frac{1}{2}V_{ijk}\D_{lj}\D_{mk}\chi_{l1}\chi_{m1},
\label{pertEOM2}\ee
which has two consequences: the first one is
\be \lambda^{(3)}_{lmp}\chi_{p1}=0,\label{con1}\ee
which can be derived from (\ref{pertEOM2}) by projecting over
$\D_l$ and it exactly reproduces (\ref{eff2vacua}) of the
effective theory method; the second consequence is
\be
\tilde{\chi}_{h2}\tilde{\D}_{hi'}(y)=-\frac{1}{2}\chi_{l1}\chi_{m1}\int
d^{D-4}y'G_{i'i}(y,y')V_{ijk}(y')\D_{lj}(y')\D_{mk}(y'),\label{con2}\ee
where $G$ still represents the Green's function for the operator
at the left hand side of (\ref{generaleigen}). Now we can write
the $\mu^3$ part of the equation (\ref{pertEOM}) as follows
\bea
&&-c_l\mu^2\chi_{l1}\D_{li}-m^2_h\tilde{\chi}_{h3}\tilde{D}_{hi}\nonumber
\\
&&-\frac{1}{2}V_{ijk}\chi_{l1}\D_{lj}\left(\tilde{\chi}_{h2}\tilde{\D}_{hk}+
\chi_{m2}\D_{mk}\right)\nonumber\\
&&-\frac{1}{2}V_{ijk}\left(\chi_{l2}\D_{lj}+
\tilde{\chi}_{h2}\tilde{\D}_{hj}\right)\chi_{m1}\D_{mk}\nonumber\\
&&-\frac{1}{3!}V_{ijkk'}\D_{lj}\D_{mk}\D_{pk'}\chi_{l1}\chi_{m1}\chi_{p1}=0.\eea
If one projects this equation over $\D_l$ and uses the equations
(\ref{con1}) and (\ref{con2}) one gets exactly the equations
(\ref{eff3vacua}). Therefore, at the order $\mu$, all the solutions
of (\ref{effvacua}) are reproduced by the D-dimensional analysis and
viceversa. Moreover we observe that these light KK modes VEVs,
predicted by the effective theory, constitute approximate solutions
of the fundamental D-dimensional EOM at leading non trivial order
because of the equation (\ref{heavyVEV}), which states that the
heavy KK modes VEVs are higher order quantity with respect to the
light KK modes VEVs.

Now we consider the mass squared eigenvalue problem which
corresponds to a solution $\Phi$; moreover we assume for simplicity
$\lambda^{(3)}_{lmp}=0$, like in the effective theory method. This
eigenvalue problem is

\be \mathcal{O}_{ij}\d\Phi_j\equiv -\partial_m\partial_m\d
\Phi_i+\frac{\partial^2V}{\partial\Phi_i\partial \Phi_j}(\Phi)\d
\Phi_j=M^2\d\Phi_i,\label{perteigen} \ee
where $\d \Phi_i$ represents the fluctuations of the scalars around
the solution $\Phi$. We observe now that the equation (\ref{perteigen}) can
be considered a time-independent Schrodinger equation:
$\mathcal{O}$ is the hamiltonian and $M^2$ the generic energy
level. Moreover we can perform a Taylor expansion of $\mathcal{O}$
around $\mu=0$:
\be \mathcal{O}=\mathcal{O}_0+\mathcal{O}_1+\mathcal{O}_2+...
\,.\ee
The operators $\mathcal{O}_1$ and $\mathcal{O}_2$ can be easily
expressed just in terms of $\chi_{l1}$ and $\chi_{l2}$ by using
(\ref{expansion}), (\ref{analitic}) and the constraints (\ref{con2})
and (\ref{heavyVEV}) which come from the EOM. From the perturbation
theory of quantum mechanics we know that the leading value of the
low energy mass spectrum is given by the eigenvalues of the
following mass squared matrix:
\be M^2_{ll'}\equiv A_{ll'}+B_{ll'},\ee
where
\be A_{ll'}\equiv <\D_l|\mathcal{O}_2|\D_{l'}>\ee
and
\be B_{ll'}\equiv
-\sum_{h}\frac{1}{m^2_h}<\D_l|\mathcal{O}_1|\tilde{\D}_h><\tilde{\D}_h|\mathcal{O}_1|\D_{l'}>.\ee
If one express the matrices $A$ and $B$ in terms\footnote{The
dependence on $\chi_{l2}$ disappears because we assume
$\lambda^{(3)}_{lmp}=0$, as one can easily check.} of $\chi_{l1}$ , one finds exactly the
corresponding result (\ref{hessian}) predicted by the effective
theory.

So we have two equivalent (at least at the leading non trivial order
in $\mu$) approaches to study the breaking of $\mathcal{G}$: the
spontaneous symmetry breaking in the 4-dimensional effective theory
and the D-dimensional analysis. We stress that, like in the simple
model of section \ref{simple4D}, also in this more general case the
heavy KK modes contribution in the effective theory can't be
neglected if one wants to reproduce the D-dimensional result, even
at the classical level. In general this is true not only in scalar
theories but also in theories which involve gauge and gravitational
interactions, as we illustrate in sections \ref{def} and \ref{u13}.

\section{A 6D Gauge and Gravitational Theory} \label{def}
\setcounter{equation}{0}

Now we consider a 6-dimensional field theory of gravity with a
$U(1)$ gauge invariance, including a charged scalar field $\phi$ and
eventually fermions. The bosonic action is\footnote{Some conventions
are fixed in appendix \ref{GR}.}
\be S_B=\int d^6 X\sqrt{-G}
\left[\frac{1}{\kappa^2}R-\frac{1}{4}F_{MN}F^{MN} -(\nabla_M
\phi)^*\nabla^M \phi -V(\phi)  \right], \label{action} \ee
where $R$ is the Ricci scalar, $\kappa$ represents the 6-dimensional
Planck scale, $F_{MN}$ is the field strength of the $U(1)$ gauge
field $A_M$, defined by
\be F_{MN}=\partial_M A_N- \partial_N A_M \ee
and
\be
\nabla_M \phi = \partial_M \phi +ie A_M \phi,\ee
where $e$ is the $U(1)$ gauge coupling. Moreover $V$ is a scalar
potential and we choose
\be V(\phi)=m^2\phi^*\phi +\xi (\phi^* \phi)^2 + \lambda,  \label{V} \ee
where $m^2$ and $\xi$ are generical real constants, with the
constraint $\xi > 0$ and $\lambda $ represents the 6-dimensional cosmological constant.

From the action (\ref{action}) we can derive the general bosonic
EOM. However we focus on the following class of backgrounds, which
are invariant under the 4-dimensional Poincar\'e group:
\bea ds^2 &=&\eta_{\mu \nu}dx^{\mu}dx^{\nu}
+g_{mn}(y)dy^m dy^n.\label{Bmetric}\\
A&=&A_m(y)dy^m, \label{BA}\\
\phi &=&\phi(y), \label{Bphi} \eea
where $g_{mn}$ is the metric of a  2-dimensional compact internal
manifold $K_2$; so the 6-dimensional space-time manifold is
$(Minkowski)_4 \times K_2$. By using (\ref{Bmetric}), (\ref{BA}) and
(\ref{Bphi}), we can write the bosonic EOM in the following form:
\bea &&\nabla^2 \phi -m^2\phi-2\xi(\phi^* \phi)\phi=0,\nonumber \\
&& \nabla_mF^{mn}+ie\left[\phi^*\nabla^n\phi
-(\nabla^n \phi)^*\phi \right]=0,\nonumber \\
&&\frac{1}{\kappa^2}R_{mn}-\frac{1}{2}F_{mp}F_n^{\,\,p}
-\frac{1}{2}(\nabla_m\phi)^*\nabla_n\phi-
\frac{1}{2}(\nabla_n\phi)^*\nabla_m\phi=0,\nonumber \\
&&\frac{1}{4}F^2-\lambda-m^2\phi^*\phi-\xi(\phi^*\phi)^2=0,
\label{EOM}
 \eea
 where $\nabla^2\equiv\nabla_m\nabla^m$ is the covariant laplacian over
the internal manifold. The equations (\ref{EOM}) must be satisfied
by the bosonic VEV.

We introduce also fermions and gauge invariant coupling
with the scalar $\phi$. In order to do that it's necessary to introduce
at least a pair of 6-dimensional Weyl spinors $\psi_+$ and $\psi_-$, where $+$ and $-$ refer here to the
6-dimensional chirality. We consider the following fermionic action:
\be S_F= \int d^6 X\sqrt{-G} \left(\overline{\psi_+}\Gamma^M \nabla_M \psi_+ +\overline{\psi_-}\Gamma^M \nabla_M \psi_-
+g_Y\phi^* \overline{\psi_+}\psi_- + g_Y\phi \overline{\psi_-}\psi_+\right), \label{fermionL}\ee
where $g_Y$ is a real Yukawa coupling constant. In (\ref{fermionL})
$\nabla_M$ represents the covariant derivative acting on spinor,
which includes the gauge and the spin connection; moreover our
conventions for $\Gamma^M$ are given in appendix \ref{GR}. The
$U(1)$ charge $e_+$ and $e_-$ of $\psi_+$ and $\psi_-$ have to
satisfy the condition $e_-=e_+ +e$ coming from the gauge invariance
of the Yukawa terms. In the following we consider the choice
$e_+=e/2$ and $e_-=3e/2$, corresponding to a simple harmonic
expansion for the compactification over $(Minkowski)_4\times S^2$.
From (\ref{fermionL}) we get the following EOM:
\be \Gamma^M \nabla_M \psi_+ +g_Y\phi^* \psi_-=0, \quad \Gamma^M \nabla_M \psi_- +g_Y\phi \psi_+=0.
\ee
Now we define the following 4-dimensional Weyl spinors:
\be \psi_{\pm L}=\frac{1-\gamma^5}{2}\psi_{\pm }, \quad \psi_{\pm R}=\frac{1+\gamma^5}{2}\psi_{\pm}, \ee
where $\gamma^5$ is the 4-dimensional chirality matrix. In terms of
$\psi_{\pm L}$ and $\psi_{\pm R}$ the EOM, for a
$(Minkowski)_4\times K_2$ background space-time, are\footnote{We
rearrange the equations in a way that the left handed and right
handed sector are split.}

\bea \left(\partial^2 + 2 \nabla_+ \nabla_- -g_Y^2|\phi|^2\right)\psi_{+L}
-\sqrt2 g_Y\left(\nabla_+\phi^* \right)\psi_{-L}=0, \nonumber \\
\left(\partial^2 + 2 \nabla_- \nabla_+ -g_Y^2|\phi|^2\right)\psi_{-L}-\sqrt2 g_Y\left(\nabla_-\phi \right)\psi_{+L}=0, \nonumber \\
\left(\partial^2 + 2 \nabla_- \nabla_+ -g_Y^2|\phi|^2\right)\psi_{+R}
+\sqrt2 g_{Y}\left(\nabla_-\phi^* \right)\psi_{-R}=0, \nonumber \\
\left(\partial^2 + 2 \nabla_+ \nabla_- -g_Y^2|\phi|^2\right)\psi_{-R}
+\sqrt2 g_{Y}\left(\nabla_+\phi \right)\psi_{+R}=0, \label{fermioneq}
\eea
where $\partial^2\equiv \eta^{\mu \nu}\partial_{\mu}\partial_{\nu}$,
\be \nabla_{\pm}= \frac{1}{\sqrt2}(\nabla_5 \pm i\nabla_6) \ee
and $\nabla_{5,6}$ are the covariant derivative components in an
orthonormal basis. The equations (\ref{fermioneq}) will be used in
order to compute the fermionic spectrum.

\subsection{The $SU(2)\times U(1)$ Background Solution} \label{su2 x u1}

An $SU(2)\times U(1)$-invariant  solution of (\ref{EOM}) is
\cite{RSS}
\bea ds^2 &=&\eta_{\mu \nu}dx^{\mu}dx^{\nu}+
a^2\left(d\theta^2+\sin^2\theta d\varphi^2\right), \label{sphere}\\
A&=&\frac{n}{2e}(\cos\theta -1)d\varphi
\equiv -\frac{n}{2e}e^3(y), \label{monopole} \\
\phi &=&0, \label{phi0} \eea
subject to the constraints
\be \lambda=\frac{n^2}{8e^2a^4}=\frac{1}{\kappa^2a^2},
\label{constraint}\ee
where $n$ is a (integer) monopole number. So here we have $K_2=S^2$,
and $a$ is the radius of $S^2$. We introduce also an orthonormal
basis in the internal cotangent space \cite{RSS}:
\be e^{\pm}(y)=\pm \frac{i}{\sqrt{2}}e^{\pm i\varphi}\left(d\theta
\pm i\sin\theta d\varphi \right). \label{epm}\ee
In the following we consider, just for simplicity, the case
\be n=2. \ee
In fact for this value of the monopole charge we can find a very
simple solution of the fundamental 6-dimensional EOMs (\ref{EOM}) which is invariant under a $U(1)$ subgroup of
$SU(2)\times U(1)$; this solution is discussed in section
\ref{u13}. Like in section \ref{general} our purpose is in fact to construct the 4-dimensional
$SU(2)\times U(1)$-invariant effective theory,
study the spontaneous symmetry breaking $SU(2)\times U(1)\rightarrow U(1)$ and the Higgs mechanism in the effective
theory and then compare the results with the corresponding quantities predicted by the 6-dimensional theory; therefore,
in order to do that, one has to find a 6-dimensional U(1)-invariant solution of the EOMs.

If $\Phi_{\lambda}$ is a field with an integer or half-integer
iso-helicity $\lambda$, we can perform an harmonic expansion
\cite{RSS}:
\be \Phi_{\lambda}(x,\theta, \phi)=\sum_{l\geq |\lambda|}
\sum_{|m|\leq l}\Phi_m^l(x)
\sqrt{\frac{2l+1}{4\pi}}\mathcal{D}^{(l)\lambda}_{m}(\theta , \varphi), \label{lexpansion}\ee
where, for a given $l$, $\mathcal{D}^{(l)\lambda}_{m}$ is a
$(2l+1)\times (2l+1)$ unitary matrix. The
$\mathcal{D}^{(l)\lambda}_{m}$ were originally introduced in
\cite{Wigner} and our conventions are given in appendix \ref{GR}.
For example $\phi$ has an expansion like (\ref{lexpansion}) with
$\lambda=1$.

The low energy 4-dimensional spectrum coming from the background
(\ref{sphere}), (\ref{monopole}) and (\ref{phi0}) is given in the
reference \cite{RSS} for the spin-1 and spin-2 sectors. The massless sector is the
following: there are a graviton (helicities
$\pm2$, $l=0$), a $U(1)$ gauge field (helicities $\pm1$, $l=0$)
coming from $\mathcal{V}_{\mu}$ and a Yang-Mills $SU(2)$ triplet (helicities $\pm 1$, $l=1$)
coming from $h_{\mu \a}$ and $\mathcal{V}_{\mu}$, where $\mathcal{V}_M$ and $h_{MN}$ are
the fluctuations of the gauge field and the metric around the solution (\ref{sphere}), (\ref{monopole}) and (\ref{phi0}).
Regarding the scalar spectrum all the scalars from $G_{MN}$ and $A_M$ have very large masses,
of the order $1/a$, and we can get only an
$SU(2)$-triplet from $\phi$ in the low energy spectrum
if we choose $m^2$ such
that
\be |\mu^2| \ll \frac{1}{a^2}, \label{assumption1}\ee
where
\be \mu^2 \equiv -\frac{1}{a^2}\eta \equiv m^2+\frac{1}{a^2}.
\label{edefin}\ee

In fact $-1/a^2$ is the eigenvalue of the laplacian operator acting
on the harmonic with $l=1$ and $\lambda=1$ , as one can check using
the related formula of \cite{RSS}. The parameter $\mu^2$ is in fact
the squared mass of the triplet from $\phi$, and it can be in
principle either positive or negative. If (\ref{assumption1}) holds
all the remaining scalars have masses at least of the order $1/a$
and they don't appear in the low energy theory. So we assume that
(\ref{assumption1}) holds. Finally in order to find the low energy
fermionic spectrum we have to calculate the associated
iso-helicities by using the explicit expression for the background
covariant derivative of $\psi_{\pm}$ along the internal space:
\be \nabla_m \psi_{\pm} = \left( \partial_m \pm\omega_m \frac{1}{2} \gamma^5 +ie_{\pm} A_m \right)\psi_{\pm}, \ee
where $\omega_\theta = 0$, $\omega_{\varphi}=\frac{i}{a}( \cos\theta
- 1)$, $e_+ = e/2$ and $e_- = 3e/2$. We get
\be \lambda_{+L}=0, \quad \lambda_{+R}=1, \quad \lambda_{-L}=2,
\quad \lambda_{-R}=1 \ee
and the corresponding expansions are given by (\ref{lexpansion}). So
the equations (\ref{fermioneq}) tell us that there are 4 zero-modes:
the $l=0$, $m=0$ mode in $\psi_{+L}$ and the $l=1$, $m=+1,-1,0$ in
$\psi_{-R}$. So we have a massless $SU(2)$ singlet from $\psi_{+L}$
and a massless $SU(2)$ triplet from $\psi_{-R}$.

\subsection{The 4D $SU(2)\times U(1)$ Effective Lagrangian \\ and the
Higgs Mechanism} \label{effective}

Now we want to study the 4D effective theory: which is the
4-dimensional
 theory obtained
from the background (\ref{sphere}), (\ref{monopole}) and (\ref{phi0}) retaining only
 the low energy spectrum we discussed at the end of subsection \ref{su2 x u1}, that is the particles with
masses much smaller than $1/a$, and integrating out all the heavy modes, namely those with mass at least of the
order $1/a$.
This is an $SU(2)\times U(1)$-invariant theory,
which includes a charged scalar, that we call $\chi$, in the $3$-dimensional representation of $SU(2)$, and, if we want, two
Weyl spinors in the $1_{1/2}$ and $3_{3/2}$ of $SU(2)\times U(1)$.
 The background (\ref{sphere}), (\ref{monopole}) and (\ref{phi0}) is the analogous of what we
 called $\Phi_{eff}$ in section \ref{general}. In this section we give
only some relevant terms\footnote{Here ``relevant terms'' has the same meaning
as in the subsection \ref{EFFGeneral}.} appearing in the lagrangian of this theory.
In particular we calculate the scalar potential, we study the Higgs mechanism,
which is active only for $\mu^2 <0$,
and we give in this case the masses of the spin-1, spin-0 and spin-1/2 particles.

Like in the general scalar theory of section \ref{general}, in the
following we perform all the calculations at the order $\eta$. If we
use the information regarding the low-energy spectrum which we
discussed at the end of subsection \ref{su2 x u1}, we can construct
some relevant terms of the 4D effective theory through the following
ansatz\footnote{The ansatz (\ref{0ansatz}) is a generalization of
the zero-mode ansatz of \cite{RSS}, which doesn't include scalar
fields.}
\bea E^a(x)&=& E^a_{\mu}(x)dx^{\mu}, \nonumber \\
E^{\alpha}(x,y)&=&e^{\alpha}(y)-\frac{\kappa}{a\sqrt{4\pi}}W_{\mu}^{\hat{\alpha}}(x)dx^{\mu}
\mathcal{D}^{\alpha}_{\hat{\alpha}}(y),
\nonumber  \\
A(x,y)&=& -\frac{n}{2ea}e^3(y) \nonumber\\
&&+\frac{1}{a\sqrt{4\pi}}V_{\mu}(x)dx^{\mu}
-\frac{n\kappa}{2ea^2\sqrt{4\pi}}U_{\mu}^{\hat{\alpha}}(x)dx^{\mu}
\mathcal{D}_{\hat{\alpha}}^3(y), \nonumber \\
 \phi(x,y)&=&\frac{1}{a}\sqrt{\frac{3}{4\pi}}\chi^{\hat{\a}}(x)\mathcal{D}_{-,\hat{\a}}(y),\nonumber \\
\psi_{+R}&=&\psi_{-L}=0, \nonumber \\
\psi_{-R}&=&\frac{1}{a}\sqrt{\frac{3}{4\pi}}\psi_R^{\hat{\a}}(x)\mathcal{D}_{-,\hat{\a}}(y),\nonumber \\
\psi_{+L}&=&\frac{1}{a\sqrt{4\pi}}\psi_L(x),
\label{0ansatz} \eea
where $E^A$, $A=0,1,2,3,+,-$, are the 6-dimensional orthonormal
1-form basis, $E^a_{\mu}$ is the 4-dimensional vielbein, $V_{\mu}$
is the 4-dimensional $U(1)$ gauge field coming from
$\mathcal{V}_{\mu}$, a linear combination\footnote{The orthogonal
linear combination has a large mass; we show this in appendix
\ref{S^2simm}.} of $W_{\mu}$ and $U_{\mu}$ is the Yang-Mills $SU(2)$
triplet \cite{RSS} coming from $h_{\mu \a}$ and $\mathcal{V}_{\mu}$;
finally $\psi_L$ and $\psi_R$ are the $SU(2)$ fermion singlet and
fermion triplet, respectively.
 Actually the ansatz (\ref{0ansatz}) is the
background (\ref{sphere}), (\ref{monopole}) and (\ref{phi0}) plus
some fluctuations, which include all the light KK states.

Now we want to write some relevant terms of the effective lagrangian
for $\chi$ by using the light-mode ansatz (\ref{0ansatz}) and by
taking into account the heavy modes contribution. The scalar
potential $\mathcal{U}$ in the 4D effective theory, including the
bilinears and the quartic interactions, is
\be \mathcal{U}(\chi)=\mu^2 \chi^{\dag} \chi
+\left(\lambda_H+c_1\lambda_G\right)\left(\chi^{\dag}\chi\right)^2
-\frac{\lambda_H+c_2\lambda_{G}}{3}\left|\chi^{\hat{\alpha}}g_{\hat{\alpha}\hat{\beta}}
\chi^{\hat{\beta}}\right|^2+...,\label{U} \ee
where $c_1$ and $c_2$ are dimensionless parameters,
\be \lambda_H\equiv \frac{9}{20 \pi a^2}\xi,\quad \quad \lambda_G\equiv \frac{9\kappa^2}{80\pi a^4} \label{lGlH}\ee
and the dots represent higher order non relevant terms, for example
terms with a product of 6 $\chi$ or 8 $\chi$. These terms don't
contribute to the VEV of $\chi$ as we want this VEV to be of the
order\footnote{The order $\eta^{1/2}$ corresponds to the order $\mu$
because of equation (\ref{edefin}).} $\eta^{1/2}$. In (\ref{U}) we
took into account that the quartic part of $\mathcal{U}$ comes from
the quartic term in the 6-dimensional potential $V$ in (\ref{V}) and
from the contribution of the heavy scalars, namely $h_{\a\b}$ and
$\mv_{\a}$, through diagrams like figure \ref{simplediagram}. The
latter contribution is represented by $c_1\lambda_G$ and
$c_2\lambda_G$, the analogous of $a_{lmpq}$ in the equation
(\ref{l4}). Moreover we give also the expression for the gauge
covariant derivative of $\chi$:
\be D_{\mu} \chi ^{\hat{\alpha}}=\partial_{\mu}\chi^{\hat{\alpha}}
+ig_1V_{\mu}\chi^{\hat{\alpha}}
+g_2\mathcal{A}_{\mu}^{\hat{\beta}}
\epsilon_{\hat{\beta}\hat{\gamma}}^{\,\,\,\,\,\,\hat{\alpha}}\chi^{\hat{\gamma}},\label{covphi}
\ee
where $\mathcal{A}_{\mu}$ is defined in appendix \ref{S^2simm} and it represents the $SU(2)$
Yang-Mills field, $\epsilon_{\hat{\gamma}\hat{\beta}\hat{\alpha}}$ is a
totally antisymmetric symbol with $\epsilon_{+-3}=i$, and
\be g_1=\frac{e}{\sqrt{4\pi}a}, \,\,\,\,
g_2=\sqrt{\frac{3}{16\pi}}\frac{\kappa}{a^2},\label{g12}\ee
are the 4-dimensional $U(1)$ and $SU(2)$ gauge couplings.

Therefore the complete lagrangian for $\chi$ is
\be \mathcal{L}_{\chi eff}=-\left(D_{\mu}\chi\right)^{\dagger}D^{\mu}\chi-\mathcal{U}(\chi).
\label{actionchi} \ee

Let's look for the points of minimum of the order $\eta^{1/2}$ of
the potential $\mathcal{U}$ in (\ref{U}). We have a minimum, in the
case $\mu^2<0$, for
\be \chi_1=\chi_2=0, \,\,\,\,
\chi_3=v\equiv\sqrt{\frac{-3\mu^2}{4\left[\lambda_H+\frac{1}{2}\left(3c_1-c_2\right)\lambda_G\right]}},\label{minimum}\ee
which corresponds to the global minimum
\be \mathcal{U}_0=0 \label{Umin}\ee
at the order $\eta$. This fact states that, at leading order, the
4-dimensional flatness condition in the background is compatible
with the procedure of the 4D effective theory. In fact
$\mathcal{U}_0$ can be interpreted as a 4-dimensional cosmological
constant and the flatness implies $\mathcal{U}_0=0$. Instead for
$\mu^2>0$ we don't have any order parameter because the global
minimum $\mathcal{U}_0=0$ corresponds to $\chi=0$.

If we take, for $\mu^2<0$,
the vacuum (\ref{minimum}),
 $SU(2)\times U(1)$ breaks to $U(1)_3$, where $U(1)_3$ is the $U(1)$-subgroup
of $SU(2)$ generated by its third generator. The gauge field of $U(1)$ and $SU(2)$ are respectively
$V_{\mu}$ and $\mathcal{A}_{\mu}$; before Higgs mechanism these gauge field are of course massless
as one can see by looking at their bilinear lagrangian given
in appendix \ref{S^2simm}. From (\ref{actionchi}) and
(\ref{covphi}) we can calculate the masses of these vector fields in the 4D effective theory after the Higgs mechanism.
We get a massless vector field $\mathcal{A}_{\mu}^3$, which corresponds
to the unbroken $U(1)_3$ gauge symmetry. Instead $V_{\mu}$ and $\mathcal{A}_{\mu}^{\pm}$ acquire respectively
the following squared masses
\be M^2_{V}  = \frac{3e^2}{8\pi
a^2}\frac{-\mu^2}{\lambda_H+\frac{1}{2}\left(3c_1-c_2\right)\lambda_G},\label{MV}
\ee
\be M^2_{V\pm}=\frac{9e^2}{16\pi
a^2}\frac{-\mu^2}{\lambda_H+\frac{1}{2}\left(3c_1-c_2\right)\lambda_G},
\label{MA} \ee
where the subscript $V$ indicates that we're dealing with vector
particles. Moreover, in the spin-0 sector, we have two physical
scalar fields: a real scalar and a complex one, which is charged
under the residual $U(1)_3$ symmetry. Their squared masses are
respectively
\be M^2_S=-2\mu^2,\ee
\be
M^2_{S\pm}=-\mu^2\frac{\lambda_H+c_2\lambda_G}{\lambda_H+\frac{1}{2}\left(3c_1-c_2\right)\lambda_G}.
\label{MS+-}\ee
Finally we can determine the fermionic spectrum by examining the
fermionic lagrangian in the effective theory:
\be \mathcal{L}_{F eff}=\overline{\psi_L}\gamma^{\mu} D_{\mu} \psi_L+\overline{\psi_R}\gamma^{\mu} D_{\mu} \psi_R
+ g_4\overline{\psi_L}\chi^{\dag}\psi_R + g_4\overline{\psi_R}\chi\psi_L,\ee
where
\be g_4=\frac{g_Y}{a\sqrt{4\pi}}. \ee
The result is a neutral Dirac fermion, with squared mass
\be M_F^2=\frac{3g_Y^2}{16\pi
a^2}\frac{-\mu^2}{\lambda_H+\frac{1}{2}\left(3c_1-c_2\right)\lambda_G},
\label{efffermion}\ee
and a pair of massless right-handed Weyl fermions.
We observe that the mass spectrum that we gave here is parametrized by the $c_i$.
Of course these constants are not free parameters but they can be in principle computed
by evaluating explicitly the heavy modes contribution. In the rest of this paper we don't compute the $c_i$ but we prove that the
4D effective theory without heavy modes contribution, that is $c_i=0$, is not correct because it predicts a wrong VEV
of the light KK scalars and a wrong mass spectrum.

\section{Symmetry Breaking in the 6D Theory} \label{u13}\setcounter{equation}{0}

Now we perform a 6-dimensional (or geometrical) analysis of
spontaneous symmetry breaking: this method corresponds to the
contents of section \ref{general} for scalar theories. Of course we
perform all the calculations at the order $\eta$, as in the
effective theory method.

The most simple solution, up to higher order terms in $\eta$, that
we find is similar to the background which appears in the reference
\cite{S} \footnote{This solution was discussed in reference
\cite{S}, but incorrectly.} :
\bea ds^2 &=&\eta_{\mu \nu}dx^{\mu}dx^{\nu}+ a^2\left[(1+|\eta |\beta \sin^{2}\theta)d\theta^2+
\sin^2\theta d\varphi^2\right], \nonumber \\
A&=& -\frac{1}{e}e^3, \nonumber \\
\phi &=&\eta^{1/2}\alpha\exp\left(i\varphi \right)
\sin\theta,
\label{solution2} \eea
where $\beta\equiv \kappa^2 |\alpha|^2$. As required, for $\eta=0$
this background reduces to the background of subsection \ref{su2 x
u1}. The value of $\phi$ in (\ref{solution2}) is proportional to the
harmonic $\D_{-,0}$, that is the harmonic with $l=1$, $\lambda=1$
and $m=0$. In order that (\ref{solution2}) is a solution, up to
$O(\eta^{3/2})$, it is necessary that (\ref{constraint}) holds and
$|\alpha |^2$ is given by the following equation:
\be \frac{1}{a^2}\eta \int \,D^* \phi + \int \,D^*L_2\phi=
2\xi\int \, D^*|\phi|^2\phi, \label{alpha}\ee
where $D$ is $\D_{-,0}$ and $L_2\phi$ is the function proportional
to $\eta^{3/2}$ in the expansion of $\nabla^2 \phi$ in powers of
$\eta^{1/2}$. Further in (\ref{alpha}) the integrals are performed
with the round $S^2$ measure. The equation (\ref{alpha}) can be
derived by putting (\ref{solution2}) in the Klein-Gordon equation.
For $\mu^2<0$ the equation (\ref{alpha}) has a solution for
\be \lambda_H>\lambda _G, \label{geometricalstability}\ee
where $\lambda_H$ and $\lambda _G$ are defined by (\ref{lGlH}),
while, for $\mu^2>0$, we have a solution for
\be \lambda_H<\lambda _G. \ee
Whether $\mu^2>0$ or $\mu^2<0$, the solution of (\ref{alpha}) is
\be |\alpha|^2=\frac{5}{|8\xi a^2-2\kappa^2|}=\frac{9}{32 \pi
a^4}\frac{1}{|\lambda_H-\lambda_G|}.\label{malpha} \ee
Note that here we have symmetry breaking for both signs of
$\mu^2$.
 This is not so interesting because the solution with
$\mu^2>0$ is unstable, as it is discussed in subsection
\ref{spin-0}. We want to stress that the value of $|\alpha|^2$
predicted by the 4D effective theory is not equal to (\ref{malpha})
if we neglect the heavy modes contribution to the effective theory,
namely for $c_i=0$: indeed in this case the effective theory
predicts a value of $|\a|^2$ equal to
\be |\alpha|_{eff}^2=\frac{9}{32 \pi
a^4}\frac{1}{\lambda_H},\label{malpha2} \ee
which is equal to (\ref{malpha}) only for $\lambda_G=0$. However
from (\ref{lGlH}) it's clear that $\lambda_G$ cannot be taken equal
to zero. Therefore we have already proved that the heavy modes
contribution is needed at least for the light mode VEV. We shall
prove that this is the case also for the mass spectrum.

As required the background (\ref{solution2}) has the symmetry
\be U(1)_3 \subset SU(2).\label{symmetrybr}\ee
  So the 4-dimensional effective low energy
theory, which follows from this background, is $U(1)_3$-invariant
and comparing these results with the effective theory predictions
makes sense.

We note that the symmetry breaking (\ref{symmetrybr}) is associated,
in the 6-dimensional theory, to a geometrical deformation of the
internal space. Further we observe that (\ref{solution2}) tell us
the heavy modes VEVs are higher order corrections with respect to
the light modes VEVs like in the scalar theories of section
\ref{general}.

Now we calculate the low energy vector, scalar and fermion spectrum
by analyzing the 4-dimensional bilinear lagrangian for the
fluctuations around the solution (\ref{solution2}).

\subsection{Spin-1 Spectrum}

The spin-1 spectrum can be calculated in a way similar to the
light mode ansatz (\ref{0ansatz}). However it must be noted that
the sectors with different $l$ no longer decouple for $\eta\neq
0$, but the mixing terms are of the order $\eta$ and they give
negligible corrections of the order $\eta^2$ to the vector boson
masses. These facts are evident from the general formula of
\cite{RS}. So we can neglect the modes with $l>1$ in the
calculation of spin-1 spectrum.
 Therefore we can compute the vector boson masses
by putting the following
ansatz in the action and integrating over the extra dimensions:

\bea E^a(x)&=& E^a_{\mu}(x)dx^{\mu}, \nonumber \\
E^{\alpha}(x,y)&=&e^{\alpha}(y,\eta)-\frac{\kappa}{a\sqrt{4\pi}}W_{\mu}^{\hat{\alpha}}(x)dx^{\mu}
\mathcal{D}^{\alpha}_{\hat{\alpha}}(y) ,
\nonumber  \\
A(x,y)&=& -\frac{1}{ea}e^3(y) \nonumber\\
&&+\frac{1}{a\sqrt{4\pi}}V_{\mu}(x)dx^{\mu}
-\frac{\kappa}{ea^2\sqrt{4\pi}}U_{\mu}^{\hat{\alpha}}(x)dx^{\mu}
\mathcal{D}_{\hat{\alpha}}^3(y), \nonumber \\
 \phi(x,y)&=&\eta^{1/2}\alpha\exp\left(i\varphi \right)
\sin\theta,
\label{ansatz} \eea
where $e^{\a}(y,\eta)$ is the orthonormal basis for the 2-dimensional metric in (\ref{solution2}):
:
\be e^{\pm}(y,\eta)=\pm \frac{i}{\sqrt{2}}e^{\pm i\varphi}
\left[\left(1+|\eta|\frac{\beta}{2}\sin^{2}\theta\right)d\theta
\pm i\sin\theta d\varphi \right]. \label{epmeps}\ee
In (\ref{ansatz}) we consider the spin-1 fluctuations but we don't
consider the spin-0 fluctuations, because they are not necessary
for the calculation of vector boson masses. It's important to note
that in (\ref{ansatz}) the VEV of $E^{\alpha}$ is
$e^{\alpha}(y,\eta)$, it's not $e^{\alpha}(y)$ as in
(\ref{0ansatz}).

From (\ref{ansatz}) it follows that some of the previous ($\eta=0$)
massless states acquire masses for $\eta\neq 0$. Up to
$O(\eta^{3/2})$, the $U(1)$ gauge boson ($l=0$) has the mass squared
\be M^2_V=\eta \frac{20}{3}\frac{e^2}{8\xi
a^2-2\kappa^2}=\frac{3e^2}{8\pi
a^2}\frac{-\mu^2}{\lambda_H-\lambda_G},\label{M0} \ee
while the Yang-Mills triplet $\mathcal{A}$ ($l=1$) is separated in
a massless gauge boson, which is associated to $U(1)_3$ gauge
invariance, and a couple of massive vector fields with the same
mass squared
\be M^2_{V\pm}=\eta \frac{10e^2}{8\xi
a^2-2\kappa^2}=\frac{9e^2}{16\pi a^2}\frac{-\mu^2}{\lambda_H-\lambda_G}.
\label{M1} \ee
By comparing (\ref{M0}) and (\ref{M1}) with (\ref{MV}) and
(\ref{MA}), we get that the heavy modes contribution is needed in
the effective theory. However we observe that the ratio
$M^2_V/M^2_{V\pm}$ is correctly predicted by the 4D effective theory
for every $c_i$.

 Since the computation of vector bosons masses is complicated we
present it explicitly. In order to prove (\ref{M0}) and (\ref{M1})
it's useful to split the action in four terms:
\be S_B=S_R+S_F+S_{\lambda}+S_{\phi}, \ee
where
\bea S_R&=&\int d^6X \sqrt{-G}\frac{1}{\kappa^2}R, \\
 S_F&=&-\frac{1}{4}\int d^6X \sqrt{-G}F^2, \\
S_{\lambda}&=&\int d^6X\sqrt{-G}\left(-\lambda \right), \\
S_{\phi}&=&\int d^6X \sqrt{-G}\left[-\left(\nabla_M\phi\right)^*\nabla^M\phi-V(\phi) \right].\label{actiondecomposition}
\eea
In appendix \ref{SFSR} we prove that the contributions coming from
$S_R$ and $S_F$ vanish, so only $S_{\phi}$ contributes to the spin-1
masses up to $O(\eta^{3/2})$. The same low-energy spin-1 masses in
(\ref{M0}) and (\ref{M1}) can be obtained also by using the general
formula of \cite{RS}, which contains all the bilinear terms in the
light cone gauge. The light cone gauge advantage is that the sectors
with different spin decouple. However the derivation that we
presented here shows that the unique contribution (at the leading
order) to the spin-1 masses comes from $S_{\phi}$, like in the
effective theory approach. This explains why the ratio
$M^2_V/M^2_{V\pm}$ is correctly predicted by the 4D effective theory
for every values of $c_i$.

\subsection{Spin-0 Spectrum} \label{spin-0}

We choose the light cone gauge \cite{RS, RSS2} in order to
evaluate the spin-0 spectrum. In this gauge we have just two
independent values for the indexes $\mu,\nu,...$ which label the
4-dimensional coordinates.
  The bilinears for the
fluctuations over the solution (\ref{solution2})
can be simply computed
with the general formula of \cite{RS}. For our model the helicity-0 $\mathcal{L}_0$ part is given by
\be \mathcal{L}_0=\mathcal{L}_0(\phi,\phi)+\mathcal{L}_0(h,h)+\mathcal{L}_0(\mathcal{V},\mathcal{V})+\mathcal{L}_0(\phi,h)
+\mathcal{L}_0(\phi,\mathcal{V})+\mathcal{L}_0(h,\mathcal{V}),\ee
where
\bea \mathcal{L}_0(\phi,\phi)&=&\phi^*\partial^2\phi+\phi^*\nabla^2\phi-\left[m^2+(4\xi+e^2)|\Phi|^2
+\kappa^2\left(\nabla_m\Phi\right)^*\nabla^m\Phi \right]|\phi|^2 \nonumber \\
&&-\frac{1}{2}\left\{\left[(2\xi-e^2)\left(\Phi^*\right)^2+\kappa^2
\left(\nabla_m\Phi \nabla^m\Phi \right)^*\right]\phi^2
+c.\,c.\right\},\label{L01} \\
\mathcal{L}_0(h,h)&=& \frac{1}{4\kappa^2}\left\{h_{mn}\partial^2
h^{mn}+h_{mn}\nabla^2 h^{mn}
+2R_{mn}^{\,\quad kl}h_l^m h_k^n \right. \nonumber \\
 &&\left.+\kappa^2h_{ks}h_{mn}F^{km}F^{sn}-2\kappa^2h^l_mh_{ln}\left[\frac{1}{2}F^m_{\,\,\,\,k}F^{nk}+\left(\nabla^m\Phi\right)^*\nabla^n\Phi
 \right]\right. \nonumber \\
&&\left.+\frac{1}{2}
h^i_i\partial^2h_j^j+\frac{1}{2}h_i^i\nabla^2h_j^j
\right\},\label{L02}\\
\mathcal{L}_0(\mv,\mv)&=& \frac{1}{2}\left\{\mv_m\partial^2\mv^m +\mv_m\nabla^2\mv^m-R_{mn}\mv^m\mv^n\right. \nonumber \\
&&\left.-2e^2|\Phi|^2\mv^m\mv_m-\kappa^2\left(F_{ml}\mv^l\right)^2
\right\}, \label{L03}\\
\mathcal{L}_0(\phi,h)&=&\nabla_lh^{lm}\phi^*\nabla_{m}\Phi+h^{mn}\left(\nabla_m
\phi \right)^*\nabla_n\Phi +c.\,c.\,,\label{L04} \\
\mathcal{L}_0(\phi,\mathcal{V})&=&2ie\mv^m\phi^*\nabla_m\Phi-\kappa^2F^{lm}\mv_m\phi^*\nabla_l\Phi+c.\,c.\,,
\label{L05}\\
\mathcal{L}_0(h,\mathcal{V})&=&\mv^n\left(\nabla_mh_{ln}F^{lm}-h_l^m\nabla_mF^l_{\,\,\,\,n}\right),
\label{L06}\eea
where $\Phi$ and $\phi$ are the background and the fluctuation of the 6-dimensional scalar.
In this vanishing-helicity sector, it turns out that we have not only mixing terms of the order $\eta$
but also mixing terms of the order $\eta^{1/2}$, coming from $\mathcal{L}_0(\phi,h)$ and $\mathcal{L}_0(\phi,\mathcal{V})$.
 So now we can't neglect
the mixing between the sectors with different values of $l$, as we
did in the helicity $\pm 1$ sector. If we integrate these
bilinears over the extra-dimensions we get an infinite dimensional
squared mass matrix. However we are interested only in the light
masses, therefore we can use the perturbation theory of quantum
mechanics in order to extract the correction of the order $\eta$
to the masses of the 6 real scalars which are massless for
$\eta=0$. We already used this method for the computation of the
mass spectrum in the scalar theories of section \ref{general}. We
explain now how to use it in this framework.

Formally we can write the bilinears $\mathcal{L}_0$ of the scalar fields in this way
\be  \mathcal{L}_0 =\frac{1}{2}S^{\dag}\partial^2 S-\frac{1}{2}S^{\dagger}\mathcal{O} S,\label{formalbilinear}\ee
where $S$ is an array which includes all the scalar fluctuations; we choose
\be S=\left(\ba {c} \phi \\
  \phi^*
 \\
  h_{++} \\
h_{--}  \\
h_{+-}  \\
  \mathcal{V}_+  \\
\mathcal{V}_-
\ea\right). \label{Sdefinition}\ee
 We have just to solve a 2-dimensional eigenvalue
problem for the squared mass operator\footnote{The matrix elements of  $\mathcal{O}$ can be computed
by comparing (\ref{formalbilinear}) with the explicit expression of $\mathcal{L}_0$.} $\mathcal{O}$:
\be \mathcal{O}S=M^2S. \ee
In particular we want to find the 6 values of $M^2$ which go to zero as $\eta $ goes to zero.
Since we're working at the order $\eta$ we decompose $\mathcal{O}$ as follows
\be
\mathcal{O}=\mathcal{O}_0+\mathcal{O}_1+\mathcal{O}_2, \label{Odecomposition}\ee
where $\mathcal{O}_0$ doesn't depend on $\eta$, $\mathcal{O}_1$ is
proportional to $\eta^{1/2}$ and $\mathcal{O}_2$ is proportional to
$\eta$. From the perturbation theory of quantum mechanics in the
degenerate case we know that the 6 values of $M^2$ we are interested
in are the eigenvalues of the following $6\times 6$
matrix\footnote{Like in section \ref{general} we use the Dirac
notation; for two states $|S_1>$ and $|S_2>$ and for an operator
$A$, $<S_1|A|S_2>$ represents $\int S_1^{\dagger}AS_2$, where the
integral is performed with the round $S^2$ metric.}:
\be M^2_{ij}=-\sum_{\tilde{i}}\frac{<i|\mathcal{O}_1|\tilde{i}><\tilde{i}|\mathcal{O}_1|j>}{M^2_{\tilde{i}}}
+<i|\mathcal{O}_2|j>,\label{QMperturbation} \ee
where $|i>$, $i=1,...6$ represent the 6 orthonormal eigenfunctions of $\mathcal{O}_0$ with vanishing eigenvalue and
they have the form
\be  |i>=\left(\ba {c} \phi \\
  \phi^*
 \\
  0 \\
.  \\
.  \\
  .  \\
0
\ea\right). \ee
Moreover $|\tilde{i}>$ are all the remaining orthonormal eigenfunctions of $\mathcal{O}_0$
and $M^2_{\tilde{i}}$ the corresponding eigenvalues. We note that the matrix elements $<i|\mathcal{O}_1|\tilde{i}>$ are
non vanishing for
\be |\tilde{i}>=\left(\ba {c} 0 \\
  0
 \\
  h_{++} \\
h_{--}  \\
h_{+-}  \\
  \mathcal{V}_+  \\
\mathcal{V}_-
\ea\right). \label{tildei}\ee
Further the operator $\mathcal{O}_1$
modifies the integration measure just by a factor proportional to
the harmonics $\mathcal{D}^{(1)}$, therefore we need just
a finite subset of ${|\tilde{i}>}$ for the evaluation of $M^2_{ij}$, namely those constructed through
the harmonics with $l=0,1,2$, which are given in
appendix \ref{GR}.
An explicit form for $|i>$ and $|\tilde{i}>$, and the
preliminary computations of the 6 eigenvalues we are interested in, are given in appendix \ref{spin-0calculation}.

We give here just the final result: we have two unphysical scalar
fields (a real and a complex one) which form the helicity-0
component of the massive vector fields; they have in fact the same
squared masses given in (\ref{M0}) and (\ref{M1}), as it's
required by Lorentz invariance, which is not manifest in the light cone gauge. Then we have a
physical real scalar and a physical complex scalar, charged under
the residual U(1) symmetry, with squared masses given respectively
by (for $\mu^2<0$)

\bea M^2_S&=&-2\mu^2, \nonumber \\
 M^2_{S\pm}&=&-\mu^2\frac{\lambda_H+\lambda_G}{\lambda_H-\lambda_G}.\label{MS} \eea
For $\mu^2>0$, we get a negative value for $M^2_S$, therefore the
corresponding solution is unstable. Note that the squared mass
$M^2_S$ has exactly the same expression as in the 4D effective
theory, for every $c_i$. But for $c_i=0$, which
corresponds to neglecting the heavy modes contribution, the
effective theory prediction for $M_{S\pm}^2$ in (\ref{MS+-}) is
not equal to the correct value (\ref{MS}).
We note that this is a physical inequivalence because the ratio $M^2_S/M^2_{S\pm}$, which is in principle
a measurable quantity, is not correctly predicted by the 4D effective theory without the heavy modes contribution. More
precisely the effective theory prediction for $M^2_S/M^2_{S\pm}$, in the case $c_i=0$, is always greater than
the correct value.

\subsection{Spin-1/2 Spectrum} \label{spin-1/2}

The spin-1/2 spectrum can be calculated by linearizing the equation of motion (\ref{fermioneq}): for $n=2$ we get
\bea &&\left(\partial^2 + 2 \nabla_+ \nabla_-
-g_Y^2|\Phi|^2\right)\psi_{+L}
=0, \nonumber \\
&&\left(\partial^2 + 2 \nabla_- \nabla_+
-g_Y^2|\Phi|^2\right)\psi_{-L}
=0, \nonumber \\
&&\left(\partial^2 + 2 \nabla_- \nabla_+
-g_Y^2|\Phi|^2\right)\psi_{+R}
+\sqrt2 g_{Y}\left(\nabla_+\Phi \right)^*\psi_{-R}=0, \nonumber \\
&&\left(\partial^2 + 2 \nabla_+ \nabla_-
-g_Y^2|\Phi|^2\right)\psi_{-R} +\sqrt2 g_{Y}\nabla_+\Phi
\,\psi_{+R}=0, \label{linearizedfermion} \eea
where $\Phi$ represents again the background of the 6-dimensional scalar, namely the third line of
(\ref{solution2}), and the covariant derivatives are evaluated with the background
metric and background gauge field given by the first and the second line of (\ref{solution2}).
These covariant derivatives are in the
$\pm$ basis defined by (\ref{epmeps}) and it includes the modified spin connection when it acts on spinors:
\be \nabla_{\a} \psi_{\pm R} = e_{\a}^m(y,\eta)\left( \partial_m \pm\omega_m \frac{1}{2} +ie_{\pm} A_m \right)\psi_{\pm R},
\ee
\be \nabla_{\a} \psi_{\pm L} = e_{\a}^m(y,\eta)\left( \partial_m \mp\omega_m \frac{1}{2}  +ie_{\pm} A_m \right)\psi_{\pm L},
 \ee
where $\omega_\theta = 0$, $\omega_{\varphi}\equiv \omega_{\varphi\,\,\, +} ^{\,\,+}$ is given in equation (\ref{spinconn})
and the value of the charges $e_{\pm}$ and the
iso-helicities\footnote{For $\eta\neq0$ we adopt the same harmonic
expansion as in the $\eta=0$ case; this gives the correct result for
the fermionic masses squared at the order $\eta$.} of the fermions
are given at the end of subsection \ref{su2 x u1}. There we give
also the fermionic massless spectrum for $\eta=0$: an $SU(2)$
singlet from $\psi_{+L}$ and an $SU(2)$ triplet from $\psi_{-R}$.

From (\ref{linearizedfermion}) it's clear that the left handed
sector doesn't present mixing terms of the order $\eta^{1/2}$ but
only of the order $\eta$. Therefore the calculation of the squared
mass $M_F^2$ of the light fermion coming from $\psi_{+L}$ is quite
easy. The result is
\be M_F^2=\frac{3g_Y^2}{16\pi a^2}\frac{-\mu^2}{\lambda_H-\lambda_G}.\label{fermionmass}\ee
Instead the evaluation of the right-handed spectrum is complicated by the presence of mixing terms of the order $\eta^{1/2}$,
as in the scalar sector. Therefore we use the perturbation theory of quantum mechanics also in the fermion right-handed sector.
Formally we can write the eigenvalue equation for the mass squared operator $\mathcal{O}$
acting in the right-handed sector as follows
\be \mathcal{O}F_R=M^2 F_R,\label{eigenfermion} \ee
where $F_R$ is an array which includes both the right-handed fermions; we choose
\be F_R=\left(\ba {c} \psi_{+R} \\
  \psi_{-R}
\ea\right). \label{FRdefinition}\ee
One can easily compute $\mathcal{O}$ acting on $F_R$ by performing
the substitution $\partial^2\rightarrow M^2$ in the last two
equations of (\ref{linearizedfermion}). Then we can proceed as in
the scalar spectrum, performing the decomposition
(\ref{Odecomposition}). However in this case the matrix $M^2_{ij}$
in (\ref{QMperturbation}) is a 3$\times$3 matrix as the number of
zero modes for $\eta=0$ in the right-handed sector is 3. Like in
the scalar spectrum we need only those $|\tilde{i}>$ vectors made
of harmonics with $l\leq 2$, because the operator $\mathcal{O}_1$
modifies the integration measure just by a factor proportional to
the harmonics $\mathcal{D}^{(1)}$.
 In appendix \ref{spin-1/2calculation} we give an expression for the $|i>,i=1,-1,0,$
vectors, for the $|\tilde{i}>$ vectors and the $M^2_{\tilde{i}}$ eigenvalues for the relevant values of $l$: $l=1,2$.
Here we give the final result: the right-handed low energy spectrum has
a pair of massless right-handed fermions as in the 4D effective theory, which have opposite charge
under the residual $U(1)$ symmetry, and a massive right-handed fermion with the same squared mass given in (\ref{fermionmass}).
This right-handed fermion together with the massive left-handed fermion form a massive Dirac spinor with mass $M_F$.

Also in the fermionic sector we note that the heavy modes
contribution is needed in order that the effective theory
reproduces the correct 6-dimensional result; this sentence is evident if one
compares the effective theory prediction (\ref{efffermion}) with
the correct result (\ref{fermionmass}).

\section{Summary and Conclusions}\label{conclusions}\setcounter{equation}{0}

The principal result of this paper is that the contribution of the
heavy KK modes to the effective 4-dimensional  action is necessary
in order to reproduce the correct D-dimensional predictions
concerning the light KK modes. We have calculated such a
contribution for a class of scalar theories. However this result
holds in a more general framework. In order to show this, we have
studied a 6-dimensional gauge and gravitational theory which
involves a complex scalar and, possibly, fermions. In particular we
have considered the compactification over $S^2$, for a particular
value of the monopole number ($n=2$), and the construction of a 4D
$SU(2)\times U(1)$ effective theory. The latter contains a scalar
triplet of  $SU(2)$ which, through an Higgs mechanism, gives masses
to the vector, scalar and fermion fields. An explicit expressions
for these masses and for the VEV of the scalar triplet was found at
the leading order in the small mass ratio $\mu/M$, where $M$ is the lightest heavy mass. On the other
hand, for $n=2$, we found a simple perturbative solution of the
fundamental 6-dimensional EOMs with the same symmetry of the 4D
effective theory in the broken phase. This solution presents a
deformation of the internal space $S^2$ to an ellipsoid, which has
isometry group $U(1)$ instead of $SU(2)$. Moreover we computed the
corresponding vector, scalar and fermion spectrum with quantum
mechanics perturbation theory technique. We have demonstrated by
direct calculation that these quantities, computed in the
6-dimensional approach, are equal to the corresponding predictions
of the 4D effective theory only if the contribution of the heavy KK
modes  are taken into account. In table \ref{comparison} we give the
spectrum predicted by the 4D effective theory for $c_i=0$, namely,
without heavy KK modes contribution, and the low energy spectrum
predicted by the 6-dimensional theory for the stable ($\mu^2<0$)
solution, that we gave in the text.
\begin{table}[top]
\begin{center}
\begin{tabular}{|l|l|l|}
\hline Squared Mass  & 4D Effective Theory &\rule{0.80cm}{0pt} 6D
Theory \quad \quad \quad  \\ \hline
 \raisebox{-0.30cm}{\rule{0pt}{0.80cm}} \quad $M^2_{V}$ & \quad \quad \quad $\frac{3e^2}{8\pi
a^2}\frac{-\mu^2}{\lambda_H}$ & \quad \quad$\frac{3e^2}{8\pi
a^2}\frac{-\mu^2}{\lambda_H-\lambda_G}$ \\ \hline
 \raisebox{-0.30cm}{\rule{0pt}{0.80cm}} \quad $M^2_{V\pm}$ & \quad \quad \quad $\frac{9e^2}{16\pi a^2}\frac{-\mu^2}{\lambda_H}$ &
\quad \quad $\frac{9e^2}{16\pi a^2}\frac{-\mu^2}{\lambda_H-\lambda_G}$ \\ \hline
\raisebox{-0.30cm}{\rule{0pt}{0.80cm}} \quad $M^2_{S}$ & \quad \quad \quad $-2\mu^2$ & \quad \quad $-2\mu^2$ \\
\hline      \raisebox{-0.30cm}{\rule{0pt}{0.80cm}} \quad $M^2_{S\pm}$ & \quad \quad \quad $-\mu^2$ &
\quad \quad $-\mu^2\frac{\lambda_H+\lambda_G}{\lambda_H-\lambda_G}$ \\\hline
\raisebox{-0.30cm}{\rule{0pt}{0.80cm}}
\quad $M^2_{F}$ & \quad \quad \quad $\frac{3g_Y^2}{16\pi a^2}\frac{-\mu^2}{\lambda_H}$
& \quad \quad $\frac{3g_Y^2}{16\pi a^2}\frac{-\mu^2}{\lambda_H-\lambda_G}$ \\
\hline  \raisebox{-0.30cm}{\rule{0pt}{0.80cm}} \quad $M^2_{F\pm}$ & \quad \quad \quad $0$ &
\quad \quad $0$ \\ \hline
\end{tabular}
\end{center}\caption{\footnotesize  The spectra predicted by the 4D effective theory without heavy modes
contribution ($c_i=0$) and by the 6D theory.} \label{comparison}
\end{table}
We observe that ratios of masses which involve only vector and
fermion excitations are correctly predicted by the 4D effective
theory even without the heavy KK modes contribution. But the ratios
of masses which involve at least one scalar mode are not correctly
predicted and the error is measured by $\lambda_G/\lambda_H$, where
$\lambda_G$ and $\lambda_H$ are defined in equations (\ref{lGlH}). We can
roughly estimate the magnitude of this disagreement: if we require
$g_1$ and $g_2$ in (\ref{g12}) to be of the order of $1$ and we
consider also the relation  between $\kappa$ and the 4-dimensional
Planck length $\kappa_4$
\be \frac{4\pi a^2}{\kappa^2}=\frac{1}{\kappa_4^2}, \ee
we get that $\sqrt {\kappa}$, $e$ and $a$ are all of the order of $\kappa_4$.
So roughly speaking the condition $\lambda_G/\lambda_H\ll 1$
becomes $\lambda_H\gg 1$, which is a strong coupling regime. Therefore we can't probably
neglect the heavy KK modes contribution and believe in the perturbation theory of quantum field theory at the same time.

Finally we note that there's a value of $c_1$ and $c_2$ ($c_1=-1/3$,
$c_2=1$) such that the effective theory VEV and vector, scalar and
fermion spectrum turn out to be correct, namely, they are equal to
the corresponding quantities given in section \ref{u13}. This is a sign of the equivalence between the
geometrical approach, which involves the deformed internal space
geometry, to the spontaneous symmetry breaking and the Higgs
mechanism in the 4D effective theory. In particular the heavy KK
modes contribution can be interpreted in a geometrical way as the
internal space deformation of the 6-dimensional solution: in fact if
we put $\b=0$ but we keep $\a \neq 0$ in (\ref{solution2}), which
corresponds to neglecting the $S^2$ deformation, we get exactly the
VEV and the spectrum predicted by the 4D effective theory without
heavy KK modes contribution.

Possible uses of our work can be its extension to the case which
resembles more the standard electro-weak theory. The latter could be
for instance the 6D gauge and gravitational theory of this paper,
compactified over $S^2$ but with monopole number $n=1$; in this case
we have in fact an Higgs doublet in the 4D effective theory. Other
interesting applications could be models without fundamental
scalars, which, in some sense, geometrize the Higgs mechanism or the
context of supersymmetric version of 6D gauge and gravitational
theories. Such supersymmetric theories have been recently
investigated in connection with attempts to find a solution to the
 cosmological dark energy  problem, a summary of which can be found in
\cite{Burgess:2004ib}.

\vspace{1cm}

{\bf Acknowledgments.} This work was supported in part by the Swiss
Science Foundation. A.S. is appreciative of the hospitality at the
IPT of Lausanne and the support by INFN.

\newpage

\appendix

{\Large \bf Appendix}

\section{Conventions and Notations} \label{GR}\setcounter{equation}{0}

We choose the signature $-,+,+,+,...$ for the metric $G_{MN}$. The Riemann
tensor is defined as follows
\be R_{MNS}^{R}=\partial_M \Gamma_{NS}^R -\partial_N \Gamma_{MS}^R +
\Gamma_{MP}^R \Gamma_{NS}^P -\Gamma_{NP}^R \Gamma_{MS}^P,  \ee
where the $\Gamma 's$ are the Levi-Civita connection. While the Ricci tensor
and the Ricci scalar
\be R_{MN}=R_{PMN}^{P}, \ \ \ \ R=G^{MN}R_{MN}.  \ee

Our choice for the 6-dimensional gamma matrices is
\be \Gamma^{\mu}=\left(\bacc 0 & \gamma^{\mu} \\
\gamma^{\mu} & 0 \ea \right), \quad  \Gamma^5=\left(\bacc 0 & \gamma^5 \\
\gamma^5 & 0 \ea \right),\quad  \Gamma^6=\left(\bacc 0 & -i \\
i & 0 \ea \right), \ee
where the $\gamma^{\mu}$ are the 4-dimensional gamma matrices and $\gamma^5$ the 4-dimensional
chirality matrix.

We define the harmonics $\mathcal{D}^{(l)\lambda}_m$ as proportional to the matrix element
\be \left<l,\lambda \right|e^{i\varphi Q_3}e^{i(\pi-\theta)Q_2}e^{i\varphi Q_3}\left|l,m\right>, \ee
where the $Q_j$, $j=1,2,3$, are the generators of $SU(2)$:
\be \left[Q_j,Q_k\right]=i\epsilon_{jkl}Q_l, \ee
where $\epsilon_{jkl}$ is the totally antisymmetric Levi-Civita symbol with $\epsilon_{123}=1$. Moreover
$\left|l,m\right>$ is the eigenvector of $\sum_j Q_j^2$ with eigenvalue $l(l+1)$ and the
eigenvector of $Q_3$ with eigenvalue $m$.

We introduce also $\mathcal{D}^{(l)}_{\lambda, m}\equiv \mathcal{D}^{(l)-\lambda}_m $ and, for $l=1$,
$\mathcal{D}_{\lambda, m}\equiv \mathcal{D}^{(1)}_{\lambda, m}$; our choice is
\be \mathcal{D}_{\hat{\alpha},\hat{\beta}}(\theta,\varphi)=
\left(\ba {ccc} \frac{1}{2}(\cos\theta+1) &
\frac{1}{2}(\cos\theta-1)e^{-2i\varphi} &
-\frac{1}{\sqrt2}\sin\theta
e^{-i\varphi} \\
 \frac{1}{2}(\cos\theta-1)e^{2i\varphi}
 & \frac{1}{2}(\cos\theta+1)
 & -\frac{1}{\sqrt2}\sin\theta
e^{i\varphi} \\
 \frac{1}{\sqrt2}\sin\theta e^{i\varphi} &
\frac{1}{\sqrt2}\sin\theta e^{-i\varphi} & \cos\theta
\ea\right).\label{harmonics} \ee
In (\ref{harmonics}) the first, second and third rows correspond
to $\hat{\alpha}=+,-,3$, the first, second and third columns to
$\hat{\beta}=+,-,3$.

While our choice for $\mathcal{D}^{(2)}_{\lambda,m}$ is
 $$\mathcal{D}^{(2)}_{\lambda,2}(\theta,\varphi)=\left(\ba {c} \frac{1}{4}\left(1+\cos\theta \right)^2 \\
  -\frac{1}{2}\sin\theta (1+\cos\theta)e^{i\varphi}
 \\
  \sqrt{\frac{3}{8}}\sin^2\theta e^{2 i\varphi} \\ -\frac{1}{2}\sin\theta (1-\cos\theta)e^{3 i\varphi}  \\
\frac{1}{4}\left( 1-\cos\theta\right)^2 e^{4i\varphi}
\ea\right),\, \mathcal{D}^{(2)}_{\lambda,1}(\theta,\varphi)=\left(\ba {c} -\frac{1}{2}\sin\theta (1+\cos\theta)e^{-i\varphi}  \\

  \frac{1}{2}(1-\cos\theta-2\cos^2\theta)  \\

 \sqrt{\frac{3}{2}}\sin\theta \cos\theta e^{i\varphi} \\

\frac{1}{4}(4\cos^2\theta-2\cos\theta -2) e^{2i\varphi}   \\

\frac{1}{2}\sin\theta (1-\cos\theta)e^{3i\varphi}
\ea\right), $$
 $$\mathcal{D}^{(2)}_{\lambda,0}(\theta,\varphi)=\left(\ba {c} \sqrt{\frac{3}{8}}\sin^2\theta e^{-2 i\varphi} \\
  \sqrt{\frac{3}{2}}\sin\theta \cos\theta e^{-i\varphi}
 \\
  \frac{1}{2}(3\cos^2\theta -1)\\-\sqrt{\frac{3}{2}}\sin\theta \cos\theta e^{i\varphi}  \\
\sqrt{\frac{3}{8}}\sin^2\theta e^{2 i\varphi}
\ea\right),\, \mathcal{D}^{(2)}_{\lambda,-1}(\theta,\varphi)=\left(\ba {c} -\frac{1}{2}\sin\theta (1-\cos\theta)e^{-3i\varphi}  \\

  \frac{1}{4}(4\cos^2\theta-2\cos\theta -2) e^{-2i\varphi}  \\

 -\sqrt{\frac{3}{2}}\sin\theta \cos\theta e^{-i\varphi} \\

\frac{1}{2}(1-\cos\theta-2\cos^2\theta)   \\

\frac{1}{2}\sin\theta (1+\cos\theta)e^{i\varphi}
\ea\right), $$
$$\mathcal{D}^{(2)}_{\lambda,-2}(\theta,\varphi)=\left(\ba {c} \frac{1}{4}\left( 1-\cos\theta\right)^2 e^{-4i\varphi} \\
  \frac{1}{2}\sin\theta (1-\cos\theta) e^{-3i\varphi}
 \\
  \sqrt{\frac{3}{8}}\sin^2\theta e^{-2 i\varphi}\\\frac{1}{2}\sin\theta (1+\cos\theta)e^{-i\varphi}   \\
\frac{1}{4}\left(1+\cos\theta \right)^2
\ea\right),$$

where $\lambda $ is the row index.

\section{Spin-1 Mass Terms from $S_F$ and $S_R$}\label{SFSR}\setcounter{equation}{0}

\subsection{ $S_F$ Contribution}\label{S_F}

In this subsection we write the contribution of
\be S_F\equiv -\frac{1}{4}\int d^6X \sqrt{-G}F^2 \ee
to the bilinear terms of $V$, $U$ and $W$.
By direct computation we get kinetic terms for $V$ and $U$ and some mass terms for $U$ and $W$:
\bea &&-\frac{1}{4}\int d^2y \e F^2=
-\frac{1}{4}V_{\mu\nu}V^{\mu\nu}K -\frac{1}{6}U_{\mu\nu}^{\hat{\alpha}}U^{\mu\nu \hb}K_{\ha \hb} \nonumber\\
     && -\frac{2}{3}U_{\mu}^{\ha}U^{\mu \hb}M^{(1)}_{\ha \hb}+
\frac{4}{3}U_{\mu}^{\ha}W^{\mu \hb}M^{(2)}_{\ha \hb}
-\frac{2}{3}W_{\mu}^{\ha}W^{\mu \hb}M^{(3)}_{\ha \hb}+...,
\eea
where the 4-dimensional curved indices $\mu$ and $\nu$ are contracted with the 4-dimensional metric $g_{\mu \nu}$,
the dots are constant terms and interaction terms, moreover
\be V_{\mu\nu}=\partial_{\mu}V_{\nu}-\partial_{\nu}V_{\mu},\,\,\,
U_{\mu\nu}^{\ha}=\partial_{\mu}U_{\nu}^{\ha}-\partial_{\nu}U_{\mu}^{\ha} \ee
and

\bea K&=&\frac{1}{4\pi a^2}\int d^2y \,\e,
\,\,\,\, K_{\ha \hb}=\frac{3}{4\pi}\left(\frac{\kappa}{\sqrt2 ea^2}\right)^2
\int d^2 y \,\e\,\D^3_{\ha}\D^3_{\hb}, \nonumber \\
M^{(1)}_{\ha \hb}&=&\frac{3}{8\pi}\left(\frac{\kappa}{\sqrt2 ea^2}\right)^2
\int d^2 y \,\e\,g^{mn}\partial_m \D_{\ha}^3\partial_n\D_{\hb}^3 \nonumber \\
M^{(2)}_{\ha \hb}&=&-\frac{3 \kappa^2}{16 \pi ea^3}\int d^2 y \,\e\,
\partial_m\D_{\ha}^3\D_{\hb}^{\alpha}e_{\alpha}^ng^{mq}F_{nq} \nonumber \\
M^{(3)}_{\ha \hb}&=&\frac{3\kappa^2}{16\pi a^2}
\int d^2 y \,\e\,\D_{\ha}^{\alpha}e_{\alpha}^m\D_{\hb}^{\beta}e_{\beta}^pF_p^{\,\,n}F_{mn}.\label{F2bil}
\eea
The results (\ref{F2bil}) are valid for all background $e^{\alpha}$ and $e^3$. We use the $SU(2)\times U(1)$ background
in the subsection \ref{S^2simm}, the $U(1)_3$ background
in the subsection \ref{S^2}.

\subsection{$S_R$ Contribution} \label{S_R}

In this subsection we write the contribution of
\be S_R\equiv \int d^6X \sqrt{-G}\frac{1}{\kappa^2}R \ee
to the bilinear terms of $W$. The complete contribution
of $S_R$ to the 4-dimensional action is given in \cite{SS} in the case of non deformed background solutions. Here we
need explicit expressions, at least for the bilinears, which are also valid for deformed solutions.
We get a kinetic term and a mass term of $W$: up to a total derivative we have

\bea &&\int d^2y \frac{1}{\kappa^2}\,\e\,R= -\frac{1}{6}W_{\mu\nu}^{\hat{\alpha}}W^{\mu\nu \hb}K'_{\ha \hb} \nonumber\\
     && +W_{\mu}^{\ha}W^{\mu \hb}M^{(4)}_{\ha \hb}+...,
\eea
where the dots include constant and interaction terms; moreover
\be W_{\mu\nu}^{\ha}=\partial_{\mu}W_{\nu}^{\ha}-\partial_{\nu}W_{\mu}^{\ha}, \ee
and
\bea K'_{\ha \hb}&=&\frac{3}{8\pi a^2}\int d^2 y \,\e\,\D^{\alpha}_{\ha}\D^{\beta}_{\hb}g_{\alpha \beta}, \nonumber \\
M^{(4)}_{\ha \hb}&=&\frac{1}{4\pi a^2}\int d^2 y \,\e\left[\partial_n\D_{\ha}^{\a}\D_{\hb}^{\b}\left(-e^m_{\a}\o_{m\,\,\,
\beta} ^{\,\,\,\c}e^n_{\c}-g_{\a\d}g^{nm}\o_{m\,\,\,
\beta} ^{\,\,\,\d}+2e^n_{\a}e^m_{\c}\o_{m\,\,\,
\beta} ^{\,\,\,\c} \right)\right.+ \nonumber \\
&&+D_{\ha}^{\a}\D_{\hb}^{\b}\left(-\frac{1}{2}\o_{n\,\,\,
\a} ^{\,\,\,\c}e^m_{\c}\o_{m\,\,\,
\beta} ^{\,\,\,\d}e^n_{\d}-\frac{1}{2}\o_{n\,\,\,
\a} ^{\,\,\,\d}g^{nm}\o_{m\d \beta}+\o_{n\,\,\,
\a} ^{\,\,\,\d}e^n_{\d}\o_{m\,\,\,
\beta} ^{\,\,\,\c}e^m_{\c}\right)\nonumber \\
&&\left.+ \partial_n\D_{\ha}^{\a}\partial_m\D_{\hb}^{\b}\left(-\frac{1}{2}e^m_{\a}e^n_{\beta}+e^n_{\a}e^m_{\beta}
-\frac{1}{2}g_{\a \b}g^{nm}\right)\right],
\label{SRbil}\eea
where $\o_{n\,\,\,\beta} ^{\,\,\,\a}$ is the 2-dimensional spin connection for $e_n^{\a}$.
The results (\ref{SRbil}) are also valid for every background $e^{\alpha}$ and $e^3$. We use the $SU(2)\times U(1)$ background
in the subsection \ref{S^2simm}, the $U(1)_3$ background
in the subsection \ref{S^2}.

\subsection{The case of $SU(2)\times U(1)$ background} \label{S^2simm}

We use now the $SU(2)\times U(1)$ background, that is $\eta =0$. This computation is
performed in \cite{RSS}. We have the following bilinear terms for $V$, $U$ and $W$:
\bea && -\frac{1}{4}V_{\mu\nu}V^{\mu\nu}
-\frac{1}{6}U_{\mu\nu}^{\hat{\alpha}}U^{\mu\nu}_{ \ha}-\frac{1}{6}W_{\mu\nu}^{\hat{\alpha}}W^{\mu\nu}_{ \ha} \nonumber\\
&&-\frac{2}{3a^2}\left(U_{\mu \hat{\alpha}}-W_{\mu
\hat{\alpha}}\right)\left(U^{\mu \hat{\alpha}}-W^{\mu
\hat{\alpha}}\right). \label{bilinearV}\eea
 If we define
\bea \mathcal{A}&=&\sqrt{\frac{1}{3}}(W+U), \nonumber \\
X&=&\sqrt{\frac{1}{3}}(W-U), \label{def2}\eea
we can write (\ref{bilinearV}) as follows
\bea
&&-\frac{1}{4}V_{\mu\nu}V^{\mu\nu}-\frac{1}{4}\mathcal{A}_{\mu\nu}^{\hat{\alpha}}\mathcal{A}^{\mu\nu}_{ \ha} \nonumber \\
&&-\frac{1}{4}X_{\mu\nu}^{\hat{\alpha}}X^{\mu\nu }_{\ha}
-\frac{2}{a^2}X_{\mu\hat{\alpha}}X^{\mu\hat{\alpha}},\eea
So $\mathcal{A}$ is a massless field, in fact it's the $SU(2)$
Yang-Mills field \cite{RSS}, while $X$ is a massive field which
can be neglected in the low-energy limit.

\subsection{The Case of $U(1)_3$ Background} \label{S^2}

Let's consider now the solution (\ref{solution2}).
 First we note that $S_R$ and $S_F$ don't give mass terms
for $V$; so the only source for the mass of $V$ is $S_{\phi}$.

We want to prove now that also the $SU(2)$
Yang-Mills fields masses don't receive contributions from $S_R$ and $S_F$.
First we give the bilinears for $U$ and $W$, which come from $S_R$ and $S_F$:
\bea &&-\frac{1}{6}U_{\mu\nu}^{\hat{\alpha}}U^{\mu\nu \hb}g_{\ha
\hb}\left(1+|\eta| \beta k_{\ha}\right)
-\frac{1}{6}W_{\mu\nu}^{\hat{\alpha}}W^{\mu\nu \hb}g_{\ha \hb}\left(1+|\eta| \beta k'_{\ha}\right) \nonumber\\
 &&-\frac{2}{3}U_{\mu}^{\ha}U^{\mu \hb}g_{\ha \hb}\left(1+|\eta| \beta m^{(1)}_{\ha}\right)
+\frac{4}{3}U_{\mu}^{\ha}W^{\mu \hb}g_{\ha \hb}\left(1+|\eta| \beta m^{(2)}_{\ha}\right) \nonumber\\
&&-\frac{2}{3}W_{\mu}^{\ha}W^{\mu \hb}g_{\ha \hb}\left(1+|\eta| \beta m^{(3)}_{\ha}\right),\label{bilinearsu13} \nonumber\\
\eea
where
\begin{displaymath} k_+=k_-=\frac{2}{5}, \,\,\, k_3=\frac{1}{5}, \,\,\,
k'_+=k'_-=\frac{3}{10}, \,\,\, k'_3=\frac{2}{5}, \nonumber\ed
\bd m^{(1)}_+=m^{(1)}_-=\frac{1}{5}, \,\,\,
m^{(1)}_3=-\frac{2}{5}, \nonumber\ed
\bd m^{(2)}_+=m^{(2)}_-=-\frac{1}{20},\,\,\,
m^{(2)}_3=-\frac{2}{5},\,\,\, m^{(3)}_+=m^{(3)}_-=-\frac{3}{10}, \,\,\,
m^{(3)}_3=-\frac{2}{5}. \nonumber\ed

In order to prove (\ref{bilinearsu13}) it's useful to use the following formula for the background
spin connection:
\be \omega_{\varphi\,\,\, +} ^{\,\,+}=-\omega_{\varphi\,\,\, -}
^{\,\,-}=\frac{i}{a}(\cos\theta-1 -\frac{1}{2}|\eta|\b\cos\theta \sin^{2}\theta) \label{spinconn}\ee
and $\omega_{\varphi\,\,\, +} ^{\,\,-}=\omega_{\varphi\,\,\, -} ^{\,\,+}=0$.

Now we define $X$ and $A$ as follows
\bea \left(1+\frac{|\eta| \beta}{2}k'_{\ha}\right)W^{\ha}=
\sqrt{\frac{3}{2}}\left(\cos\theta_{\eta}^{\ha}X^{\ha}+\sin\theta_{\eta}^{\ha}\mathcal{A}^{\ha}\right), \nonumber\\
\left(1+\frac{|\eta| \beta}{2}k_{\ha}\right)U^{\ha}=
\sqrt{\frac{3}{2}}\left(-\sin\theta_{\eta}^{\ha}X^{\ha}+\cos\theta_{\eta}^{\ha}\mathcal{A}^{\ha}\right),
\label{defeps} \eea
where the angle $\theta_{\eta}^{\ha}$ is defined by
\be \cos\theta_{\eta}^{\ha}=\frac{1+|\eta| \beta
\delta^{\ha}}{\sqrt2},\,\,\,
\sin\theta_{\eta}^{\ha}=\frac{1-|\eta| \beta
\delta^{\ha}}{\sqrt2},\ee
and the quantities $\d^{\ha}$ are not still fixed. It's simple to check that the kinetic terms for $X$ and $\mathcal{A}$ are in the
standard form for every $\d^{\ha}$ up to $O(\eta^{3/2})$.
The definition (\ref{defeps}) reduce to (\ref{def2}) for $\eta=0$.

If we choose
\be \d^{\ha}=\frac{1}{8}\left(m^{(3)}_{\ha}-k'_{\ha}-m^{(1)}_{\ha}+k_{\ha}\right) \ee
we have no mass terms for $\mathcal{A}$ coming from $S_R+S_F$.

So the only source for the spin-1 low energy spectrum is $S_{\phi}$ and the
result is given in equations (\ref{M0}) and (\ref{M1}).

\section{Explicit Calculation of Spin-0 Spectrum for the 6D Theory}\label{spin-0calculation}\setcounter{equation}{0}

As we pointed out in the text, in order to find the spin-0 spectrum
the expression of the $|i>$, $i=1,...,6$, vectors is needed; these
are defined by $\mathcal{O}_0|i>=0$, which is equivalent to
$\nabla^2\phi+\phi/a^2=0$, where $\nabla^2\phi$ is the laplacian
over the charged scalar $\phi$, calculated with the round $S^2$
metric. Our choice for the orthonormal vectors\footnote{We express a
generic vector as in (\ref{Sdefinition}).}  $|i>$ is

 $$|1>=\frac{1}{\sqrt{2}}\left(\ba {c} \sqrt{\frac{3}{4\pi}}\mathcal{D}^{(1)}_{-1,1}\\
  \sqrt{\frac{3}{4\pi}}\left(\mathcal{D}^{(1)}_{-1,1}\right)^*
 \\
  0  \\
.\\
. \\
 .  \\

 0
\ea\right),\,|2>=\frac{1}{\sqrt{2}}\left(\ba {c} i\sqrt{\frac{3}{4\pi}}\mathcal{D}^{(1)}_{-1,1}\\
  -i\sqrt{\frac{3}{4\pi}}\left(\mathcal{D}^{(1)}_{-1,1}\right)^*
 \\
  0  \\
.\\
. \\
 .  \\
 0
\ea\right), $$
 $$|3>=\frac{1}{\sqrt{2}}\left(\ba {c} \sqrt{\frac{3}{4\pi}}\mathcal{D}^{(1)}_{-1,0}\\
  \sqrt{\frac{3}{4\pi}}\left(\mathcal{D}^{(1)}_{-1,0}\right)^*
 \\
  0  \\
.\\
. \\
 .  \\

 0
\ea\right), \,|4>=\frac{1}{\sqrt{2}}\left(\ba {c} i\sqrt{\frac{3}{4\pi}}\mathcal{D}^{(1)}_{-1,0}\\
  -i\sqrt{\frac{3}{4\pi}}\left(\mathcal{D}^{(1)}_{-1,0}\right)^*
 \\
  0  \\
.\\
. \\
 .  \\
 0
\ea\right), $$
$$|5>=\frac{1}{\sqrt{2}}\left(\ba {c} \sqrt{\frac{3}{4\pi}}\mathcal{D}^{(1)}_{-1,-1}\\
  \sqrt{\frac{3}{4\pi}}\left(\mathcal{D}^{(1)}_{-1,-1}\right)^*
 \\
  0  \\
.\\
. \\
 .  \\

 0
\ea\right), \,|6>=\frac{1}{\sqrt{2}}\left(\ba {c} i\sqrt{\frac{3}{4\pi}}\mathcal{D}^{(1)}_{-1,-1}\\
  -i\sqrt{\frac{3}{4\pi}}\left(\mathcal{D}^{(1)}_{-1,-1}\right)^*
 \\
  0  \\
.\\
. \\
 .  \\
 0
\ea\right). $$

Another ingredient for the calculation of the spin-0 spectrum is an
explicit expression of the vectors $|\tilde{i}>$ and of the
eigenvalues $M^2_{\tilde{i}}$. As we explained in the text, only the
$|\tilde{i}>$ like (\ref{tildei}) and made of $l=0,1,2$ harmonics
are needed. The $|\tilde{i}> $ vectors must satisfy the following
eigenvalue equations\footnote{We derive (\ref{eigen}) evaluating
(\ref{L02}), (\ref{L03}) and (\ref{L06}) in the basis (\ref{epm})
and performing the redefinition $h_{\pm \pm}\rightarrow
\sqrt{2}\kappa h_{\pm \pm}$ and $h_{+-}\rightarrow
h_{+-}\kappa/\sqrt{2}$, which normalizes the kinetic terms in the
standard way.}:
\bea &&-\nabla^2h_{++}+2R_{+-+-}h_{++}-2\kappa^2F_{+-}^2h_{++}-\sqrt{2}\kappa\nabla_{+}\mv_{+}F_{-+}=M^2h_{++},\nonumber \\
&&-\nabla^2h_{--}+2R_{+-+-}h_{--}-2\kappa^2F_{+-}^2h_{--}+\sqrt{2}\kappa\nabla_{-}\mv_{-}F_{-+}=M^2h_{--},\nonumber \\
&&-\nabla^2h_{+-}-R_{+-+-}h_{+-}-\frac{\kappa}{\sqrt2}\nabla_+\mv_-F_{-+}+
\frac{\kappa}{\sqrt2}\nabla_-\mv_+F_{-+}=M^2h_{+-},\nonumber \\
&&-\nabla^2\mv_{+}+R_{+-}\mv_{+}-\kappa^2\mv_+ F^2_{+-}+
\frac{\kappa}{\sqrt2}\nabla_+h_{+-}F_{-+}-\sqrt2 \kappa\nabla_-h_{++}F_{-+}=M^2\mv_{+},\nonumber \\
&&-\nabla^2\mv_{-}+R_{+-}\mv_{-}-\kappa^2\mv_- F^2_{+-}-
\frac{\kappa}{\sqrt2}\nabla_-h_{+-}F_{-+}+\sqrt2 \kappa\nabla_+h_{--}F_{-+}=M^2\mv_{-},\nonumber \\
\label{eigen}\eea
where the background objects ($\nabla^2$, $R_{+-+-}$,...) correspond
to the background (\ref{sphere}), (\ref{monopole}) and (\ref{phi0}).
We can transform the differential problem (\ref{eigen}) into an
algebraic one by using the expansion (\ref{lexpansion}). We get an
eigenvalue problem for every value of $l$ and we give now an
explicit expression for the $|\tilde{i}>$ vectors for the relevant
value of $l$, namely $l=0,1,2$. For $l=0$ we get just one
eigenvector $|\tilde{1}>$ with $M^2=1/a^2$:
\be |\tilde{1}>=\left(\ba {c} 0 \\
  0
 \\
  0 \\
0  \\
1/\sqrt{4\pi}  \\
  0  \\
0
\ea\right). \ee
For $l=1$ we get three different eigenvalues: $M^2=2/a^2, 4/a^2, 5/a^2$ . The eigenvectors which
correspond to $M^2=2/a^2$ are
\be |\tilde{2}_0>, \quad \frac{1}{\sqrt2}\left(|\tilde{2}_{1}>+|\tilde{2}_{-1}>\right),\quad
\frac{1}{\sqrt2 i}\left(|\tilde{2}_{1}>-|\tilde{2}_{-1}>\right), \ee
where
\be |\tilde{2}_{m}>\equiv\frac{1}{\sqrt6}\left(\ba {c} 0 \\
  0
 \\
  0 \\
0  \\
2\sqrt{\frac{3}{4\pi}}\mathcal{D}^{(1)}_{0,m}  \\
  -\sqrt{\frac{3}{4\pi}}\mathcal{D}^{(1)}_{1,m}  \\
-\sqrt{\frac{3}{4\pi}}\mathcal{D}^{(1)}_{-1,m}
\ea\right). \ee
Instead the eigenvectors which
correspond to $M^2=4/a^2$ are
\be i|\tilde{3}_0>, \quad \frac{1}{\sqrt2 i}\left(|\tilde{3}_{1}>+|\tilde{3}_{-1}>\right),\quad
\frac{1}{\sqrt2 }\left(|\tilde{3}_{1}>-|\tilde{3}_{-1}>\right), \ee
where
\be |\tilde{3}_{m}>\equiv\frac{1}{\sqrt2}\left(\ba {c} 0 \\
  0
 \\
  0 \\
0  \\
0  \\
  -\sqrt{\frac{3}{4\pi}}\mathcal{D}^{(1)}_{1,m}  \\
\sqrt{\frac{3}{4\pi}}\mathcal{D}^{(1)}_{-1,m}
\ea\right). \ee
Moreover the eigenvectors which
correspond to $M^2=5/a^2$ are
\be |\tilde{4}_0>, \quad \frac{1}{\sqrt2 }\left(|\tilde{4}_{1}>+|\tilde{4}_{-1}>\right),\quad
\frac{1}{\sqrt2 i}\left(|\tilde{4}_{1}>-|\tilde{4}_{-1}>\right), \ee
where
\be |\tilde{4}_{m}>\equiv\frac{1}{\sqrt3}\left(\ba {c} 0 \\
  0
 \\
  0 \\
0  \\
 \sqrt{\frac{3}{4\pi}}\mathcal{D}^{(1)}_{0,m} \\
  \sqrt{\frac{3}{4\pi}}\mathcal{D}^{(1)}_{1,m}  \\
\sqrt{\frac{3}{4\pi}}\mathcal{D}^{(1)}_{-1,m}
\ea\right). \ee
Finally, for $l=2$ the values of $M^2$ are given by
\be a^2M^2=6,\,\,2(3-\sqrt3),\,\, 2(3+\sqrt3),\,\,\frac{1}{2}(13-\sqrt{73}) ,\,\,\frac{1}{2}(13+\sqrt{73}).\ee
The eigenvectors with $a^2M^2=6$ are
\bea &&|\tilde{5}_0>,\quad \frac{1}{\sqrt2}\left(|\tilde{5}_{1}>-|\tilde{5}_{-1}>\right),\quad
\frac{1}{\sqrt2i}\left(|\tilde{5}_{1}>+|\tilde{5}_{-1}>\right),\nonumber \\
&&\frac{1}{\sqrt2}\left(|\tilde{5}_{2}>+|\tilde{5}_{-2}>\right),\quad
\frac{1}{\sqrt2i}\left(|\tilde{5}_{2}>-|\tilde{5}_{-2}>\right),
\eea
where
\be |\tilde{5}_{m}>\equiv\frac{1}{3\sqrt2}\left(\ba {c} 0 \\
0 \\
-\sqrt2 \sqrt{\frac{5}{4\pi}}\mathcal{D}^{(2)}_{2,m} \\
  -\sqrt2 \sqrt{\frac{5}{4\pi}}\mathcal{D}^{(2)}_{-2,m}
 \\
  -2\sqrt3 \sqrt{\frac{5}{4\pi}}\mathcal{D}^{(2)}_{0,m} \\
  -\sqrt{\frac{5}{4\pi}}\mathcal{D}^{(2)}_{1,m}  \\
\sqrt{\frac{5}{4\pi}}\mathcal{D}^{(2)}_{-1,m}
\ea\right). \ee
For $a^2M^2=2(3-\sqrt3)$ we have the eigenvectors
\bea &&i|\tilde{6}_0>,\quad \frac{1}{\sqrt2}\left(|\tilde{6}_{1}>+|\tilde{6}_{-1}>\right),\quad
\frac{1}{\sqrt2i}\left(|\tilde{6}_{1}>-|\tilde{6}_{-1}>\right),\nonumber \\
&&\frac{1}{\sqrt2}\left(|\tilde{6}_{2}>-|\tilde{6}_{-2}>\right),\quad
\frac{1}{\sqrt2i}\left(|\tilde{6}_{2}>+|\tilde{6}_{-2}>\right),
\eea
where
\be |\tilde{6}_{m}>\equiv\frac{1}{\sqrt{2(3+\sqrt3)}}\left(\ba {c} 0 \\
0 \\
-\frac{1+\sqrt3}{\sqrt2} \sqrt{\frac{5}{4\pi}}\mathcal{D}^{(2)}_{2,m} \\
 \frac{1+\sqrt3}{\sqrt2} \sqrt{\frac{5}{4\pi}}\mathcal{D}^{(2)}_{-2,m}
 \\
 0 \\
  \sqrt{\frac{5}{4\pi}}\mathcal{D}^{(2)}_{1,m}  \\
\sqrt{\frac{5}{4\pi}}\mathcal{D}^{(2)}_{-1,m}
\ea\right). \ee
For $a^2M^2=2(3+\sqrt3)$ we have the eigenvectors
\bea &&i|\tilde{7}_0>,\quad \frac{1}{\sqrt2}\left(|\tilde{7}_{1}>+|\tilde{7}_{-1}>\right),\quad
\frac{1}{\sqrt2i}\left(|\tilde{7}_{1}>-|\tilde{7}_{-1}>\right),\nonumber \\
&&\frac{1}{\sqrt2}\left(|\tilde{7}_{2}>-|\tilde{7}_{-2}>\right),\quad
\frac{1}{\sqrt2i}\left(|\tilde{7}_{2}>+|\tilde{7}_{-2}>\right),
\eea
where
\be |\tilde{7}_{m}>\equiv\frac{1}{\sqrt{2(3-\sqrt3)}}\left(\ba {c} 0 \\
0 \\
-\frac{1-\sqrt3}{\sqrt2} \sqrt{\frac{5}{4\pi}}\mathcal{D}^{(2)}_{2,m} \\
 \frac{1-\sqrt3}{\sqrt2} \sqrt{\frac{5}{4\pi}}\mathcal{D}^{(2)}_{-2,m}
 \\
 0 \\
  \sqrt{\frac{5}{4\pi}}\mathcal{D}^{(2)}_{1,m}  \\
\sqrt{\frac{5}{4\pi}}\mathcal{D}^{(2)}_{-1,m}
\ea\right). \ee
Then for $a^2M^2=(13-\sqrt{73})/2$:
\bea &&|\tilde{8}_0>,\quad \frac{1}{\sqrt2}\left(|\tilde{8}_{1}>-|\tilde{8}_{-1}>\right),\quad
\frac{1}{\sqrt2i}\left(|\tilde{8}_{1}>+|\tilde{8}_{-1}>\right),\nonumber \\
&&\frac{1}{\sqrt2}\left(|\tilde{8}_{2}>+|\tilde{8}_{-2}>\right),\quad
\frac{1}{\sqrt2i}\left(|\tilde{8}_{2}>-|\tilde{8}_{-2}>\right),
\eea
where
\be |\tilde{8}_{m}>\equiv\frac{1+\sqrt{73}}{\sqrt{438+30\sqrt{73}}}\left(\ba {c} 0 \\
0 \\
\frac{13\sqrt2+\sqrt{146}}{2(1+\sqrt{73})} \sqrt{\frac{5}{4\pi}}\mathcal{D}^{(2)}_{2,m} \\
  \frac{13\sqrt2+\sqrt{146}}{2(1+\sqrt{73})} \sqrt{\frac{5}{4\pi}}\mathcal{D}^{(2)}_{-2,m}
 \\
  -\frac{4\sqrt3}{1+\sqrt{73}} \sqrt{\frac{5}{4\pi}}\mathcal{D}^{(2)}_{0,m} \\
  -\sqrt{\frac{5}{4\pi}}\mathcal{D}^{(2)}_{1,m}  \\
\sqrt{\frac{5}{4\pi}}\mathcal{D}^{(2)}_{-1,m}
\ea\right). \ee
Finally for $a^2M^2=(13+\sqrt{73})/2$:
\bea &&|\tilde{9}_0>,\quad \frac{1}{\sqrt2}\left(|\tilde{9}_{1}>-|\tilde{9}_{-1}>\right),\quad
\frac{1}{\sqrt2i}\left(|\tilde{9}_{1}>+|\tilde{9}_{-1}>\right),\nonumber \\
&&\frac{1}{\sqrt2}\left(|\tilde{9}_{2}>+|\tilde{9}_{-2}>\right),\quad
\frac{1}{\sqrt2i}\left(|\tilde{9}_{2}>-|\tilde{9}_{-2}>\right),
\eea
where
\be |\tilde{9}_{m}>\equiv\frac{1-\sqrt{73}}{\sqrt{438-30\sqrt{73}}}\left(\ba {c} 0 \\
0 \\
\frac{13\sqrt2-\sqrt{146}}{2(1-\sqrt{73})} \sqrt{\frac{5}{4\pi}}\mathcal{D}^{(2)}_{2,m} \\
  \frac{13\sqrt2-\sqrt{146}}{2(1-\sqrt{73})} \sqrt{\frac{5}{4\pi}}\mathcal{D}^{(2)}_{-2,m}
 \\
  -\frac{4\sqrt3}{1-\sqrt{73}} \sqrt{\frac{5}{4\pi}}\mathcal{D}^{(2)}_{0,m} \\
  -\sqrt{\frac{5}{4\pi}}\mathcal{D}^{(2)}_{1,m}  \\
\sqrt{\frac{5}{4\pi}}\mathcal{D}^{(2)}_{-1,m}
\ea\right). \ee

We can now calculate the $6\times6$ matrix $M^2_{ij}$ given in (\ref{QMperturbation}). In order to do that
we need just the matrix elements $<i|\mathcal{O}_1|\tilde{i}>$ and $<i|\mathcal{O}_2|j>$, which can be
computed by evaluating\footnote{For the background solution (\ref{sphere}), (\ref{monopole}) and (\ref{phi0})
we have  $\mathcal{L}_0(\phi,\mv)=0$.} $\mathcal{L}_0(\phi,h)$ and $\mathcal{L}_0(\phi,\phi)$, which appears
in (\ref{L01}) and (\ref{L04}), in the $\pm$ basis given in (\ref{epm}). After the redefinitions
$h_{\pm \pm}\rightarrow \sqrt{2}k h_{\pm \pm}$ and $h_{+-}\rightarrow h_{+-}k/\sqrt{2}$,
which normalize the kinetic terms in the standard way, we get (for $n=2$)
\bea
\mathcal{L}_0(\phi,h)&=&\sqrt2 \kappa\nabla_{+}\Phi\nabla_+h_{--}\phi^{*}
+\frac{\kappa}{\sqrt2}\nabla_{+}\Phi\nabla_-h_{+-}\phi^{*} \nonumber \\
&&+\sqrt2 \kappa\nabla_{+}\Phi h_{--}\left(\nabla_-\phi\right)^{*}
+\frac{\kappa}{\sqrt2}\nabla_{+}\Phi
h_{+-}\left(\nabla_+\phi\right)^{*}+c.c. \, ,\nonumber \\
\mathcal{L}_0(\phi,\phi)&=&\phi^*\partial^2\phi-\phi^*\left[-\nabla^2+m^2+(e^2+4\xi)\Phi^*
\Phi
+\kappa^2\nabla_+\Phi \left(\nabla_+\Phi\right)^*\right]\phi \nonumber \\
&&-\frac{1}{2}\left[\phi(2\xi-e^2)\left(\Phi^*\right)^2\phi+c.c.
\right].\eea
By using these expressions and the values of $|i>$ and $|\tilde{i}>$ given before, we find the following expression
for $M^2_{ij}$:
\be  \{M^2_{ij}\}=
\left(\ba {cccccc} a_1 &
0 &
0 & 0 & a_4 & 0 \\
 0
 & a_1
 & 0 & 0 & 0 & -a_4 \\
0 & 0 & a_2 & 0 & 0 & 0 \\
0 & 0 & 0 & a_3 & 0 & 0 \\
a_4 &
0 &
0 & 0 & a_1 & 0 \\
0
 & -a_4
 & 0 & 0 & 0 & a_1 \ea\right),\label{M2ij} \ee
where
\bea a_1&=&\frac{|\eta|}{a^2}\left(-sign(\eta) +\frac{3}{10}\b + \frac{12}{5}\frac{\b \xi a^2}{\kappa^2}\right), \nonumber \\
 a_2&=&\frac{|\eta|}{a^2}\left(-sign(\eta) -\frac{6}{5}\b + \frac{24}{5}\frac{\b \xi a^2}{\kappa^2}\right), \nonumber \\
a_3&=&\frac{|\eta|}{a^2}\left(-sign(\eta) +\frac{4}{15}\b + \frac{8}{5}\frac{\b \xi a^2}{\kappa^2}\right), \nonumber \\
a_4&=&\frac{|\eta|}{a^2}\b\left(\frac{3}{10}- \frac{4}{5}\frac{ \xi a^2}{\kappa^2}\right).\eea
By diagonalizing $M^2_{ij}$, we found exactly the spectrum that we discussed in the subsection \ref{spin-0}: the squared
masses of the vector particles are reproduced\footnote{In order to see that we use the background
constraints (\ref{constraint}).}, as required by the light cone gauge; moreover we get the two masses squared given in
(\ref{MS}).

\section{Explicit Calculation of Spin-1/2 Spectrum for the 6D Theory}\label{spin-1/2calculation}\setcounter{equation}{0}

Here we concentrate on the right-handed sector, which is the non trivial one
because it presents $\eta^{1/2}$ mixing terms.

The eigenvalue equations for the unperturbed ($\eta=0$) mass squared operator $\mathcal{O}_0$,
acting on the right-handed sector, are
\bea - 2 \nabla_- \nabla_+\psi_{+R}
=M^2\psi_{+R}, \nonumber \\
- 2\nabla_+ \nabla_- \psi_{-R}
=M^2\psi_{-R}, \label{unperturbedfermion}
\eea
The differential equation (\ref{unperturbedfermion}) can be
transformed in an algebraic one through the harmonic expansion,
remembering the iso-helicities of $\psi_{+R}$ and $\psi_{-R}$:
$\lambda_{+R}=\lambda_{-R}=1$.
 Therefore an explicit expression for the vectors $|i>$, which satisfies by definition $\mathcal{O}_0|i>=0$, is given by
\be |i>=\left(\ba {c} 0 \\
  \sqrt{\frac{3}{4\pi}}\mathcal{D}^{(1)}_{-1,i}
\ea\right),\quad \quad i=1,-1,0. \ee
We give also an expression for the vectors $|\tilde{i}>$ and the corresponding non vanishing eigenvalues $M^2_{\tilde{i}}$.
For $l=1$, we have just one eigenvalue $M^2=2/a^2$ and the corresponding eigenvectors are
\be |\tilde{1}_m>=\left(\ba {c}
  \sqrt{\frac{3}{4\pi}}\mathcal{D}^{(1)}_{-1,m} \\
0
\ea\right). \ee
For $l=2$ we have an eigenvalue $M^2=6/a^2$, which corresponds to the eigenvectors
\be |\tilde{2}_m>=\left(\ba {c}
  \sqrt{\frac{5}{4\pi}}\mathcal{D}^{(2)}_{-1,m} \\
0
\ea\right), \ee
and an eigenvalue $M^2=4/a^2$, which corresponds to the eigenvectors
\be |\tilde{3}_m>=\left(\ba {c} 0 \\
  \sqrt{\frac{5}{4\pi}}\mathcal{D}^{(1)}_{-1,m}
\ea\right). \ee
By inserting these eigenvectors and eigenvalues in the expression (\ref{QMperturbation}) we get
\be M^2_{ij}=diag\left(\,\,0,\,\,0,\,\,\frac{2}{3}|\eta|g^2_Y\frac{\b}{\kappa^2}\right), \ee
which corresponds to the spectrum we discussed at the end of section
\ref{spin-1/2}.

\end{document}